\documentclass[manuscript,nonacm]{acmart}
\AtBeginDocument{%
  }

\setcopyright{acmlicensed}
\copyrightyear{2018}
\acmYear{2018}
\acmDOI{XXXXXXX.XXXXXXX}
\acmConference[Conference acronym 'XX]{Make sure to enter the correct
  conference title from your rights confirmation email}{June 03--05,
  2018}{Woodstock, NY}
\acmISBN{978-1-4503-XXXX-X/2018/06}

\usepackage{amsmath}
\usepackage{subcaption}
\usepackage{framed}
\usepackage{enumitem}
\usepackage{multirow}
\usepackage{multicol}
\usepackage{booktabs}
\usepackage{makecell}
\usepackage{tabularx}

\usepackage{amssymb}
\usepackage{amsfonts}
\usepackage{threeparttable}
\usepackage{wrapfig}
\usepackage[table]{xcolor}
\definecolor{correctbg}{RGB}{206,234,180}
\definecolor{badutilitybg}{RGB}{255,255,150}
\definecolor{notdpbg}{RGB}{255,200,194}

\usepackage{float}
\usepackage{placeins}
\usepackage{listings}
\usepackage{xcolor}
\begin{document}

\newcommand{\name}{DPCheatSheet}
\title{\name: Using Worked and Erroneous LLM-usage Examples to Scaffold Differential Privacy Implementation}
\newcommand{\haojian}[1]{\textcolor{violet}{#1}}
\newcommand{\placeholder}[1]{\textcolor{red}{#1}}
\newcommand{\sssec}[1]{\vspace*{0.05in}\noindent\textbf{#1}}

\author{Shao-Yu Chu}
\email{shaoyuchu@ucsd.edu}
\orcid{0009-0000-8487-8678}
\affiliation{%
  \institution{University of California, San Diego}
  \state{California}
  \country{USA}
}

\author{Yuhe Tian}
\email{yut009@ucsd.edu}
\orcid{0009-0008-5233-3647}
\affiliation{%
  \institution{University of California, San Diego}
  \state{California}
  \country{USA}
}

\author{Yu-Xiang Wang}
\email{yuxiangw@ucsd.edu}
\orcid{0000-0002-6403-212X}
\affiliation{%
  \institution{University of California, San Diego}
  \state{California}
  \country{USA}
}
\author{Haojian Jin}
\email{haojian@ucsd.edu}
\orcid{0000-0001-5212-2235}
\affiliation{%
  \institution{University of California, San Diego}
  \state{California}
  \country{USA}
}

\renewcommand{\shortauthors}{Chu et al.}

\begin{abstract}

This paper explores how programmers without specialized expertise in differential privacy (DP) (i.e., novices) can leverage LLMs to implement DP programs with minimal training. We first conducted a need-finding study with 6 novices and 3 experts to understand how they utilize LLMs in DP implementation. While DP experts can implement correct DP analyses through a few prompts, novices struggle to articulate their requirements in prompts and lack the skills to verify the correctness of the generated code. We then developed \name, an instructional tool that helps novices implement DP using LLMs. \name\ combines two learning concepts: it annotates an expert's workflow with LLMs as a worked example to bridge the expert mindset to novices, and it presents five common mistakes in LLM-based DP code generation as erroneous examples to support error-driven learning. We demonstrated the effectiveness of \name\ with an error identification study and an open-ended DP implementation study.
\end{abstract}

\ccsdesc[300]{Security and privacy~Usability in security and privacy}
\ccsdesc[300]{Security and privacy~Privacy-preserving protocols}

\keywords{Differential Privacy, Large Language Models, Instructional Tools, Example-Based Learning}

\received{20 February 2007}
\received[revised]{12 March 2009}
\received[accepted]{5 June 2009}

\maketitle

\section{Introduction}

Differential privacy (DP) has emerged as the leading concept for ensuring privacy in statistical data analysis. Yet, real-world deployments remain few ~\cite{Cummings2024Advancing, garrido2022lessons} and mostly confined to large organizations~\cite{desfontainesblog20211001} such as the US Census Bureau~\cite{abowd2018us} and Apple~\cite{erlingsson2014rappor}. One major barrier for individuals or startups in protecting sensitive data with DP is its steep learning curve~\cite{ngong2024evaluating, sarathy2023don, dibia2024sok}.

Researchers have explored several approaches to ease the learning process of differential privacy, including providing public tutorials~\cite{near_abuah_2021, cowan2024hands}, enhancing differential privacy libraries~\cite{gaboardi2020programming, pydp, berghel2022tumult}, and incorporating the topic into university courses~\cite{cs860kamath, cs3110near, dsc291wang, cs2080Vadhan}.  Most of the approaches assume a conventional learning paradigm, where developers first search for learning materials, learn the relevant concepts, and then manually implement the solutions. However, we observed a generational shift in our recent differential privacy course, where students ask LLMs (e.g., ChatGPT) to generate solutions for them, outsourcing problem-solving to the model from the outset~\cite{stackoverflow_2025_developer_survey_ai, klemmer2024using}.

This paper explores how to better support novices in implementing differential privacy (DP) solutions with the help of LLMs. We focus on self-directed learning in DP programming, recognizing that over 80\% of developers acquire skills through self-study rather than formal education~\cite{stackoverflow_2025_learning_to_code}. Many organizations, especially startups, lack the resources to hire dedicated DP specialists, making it necessary for developers to build expertise on their own~\cite{shafieinejad2025adopt, almeida2003startup}.
Consider a software developer at a startup who wants to write a differentially private program to analyze their data. They have solid programming experience and a basic grasp of statistics, but only limited exposure to DP---perhaps from watching a few introductory videos. Our approach aims to lower the barrier to implementing DP, thereby enabling broader DP deployment in practice.

Our exploration focuses on two key research questions:

\noindent\textbf{RQ1: Can developers use LLMs to implement differential privacy solutions? If not, what are the key barriers?}

We first conducted a needfinding study~\cite{patnaik1999needfinding} with 6 novices and 3 experts to understand how they utilize LLMs in DP implementation (Section~\ref{need-finding}). We asked all participants to implement a differentially private z-score analysis~\cite{abdi2007z} with the help of ChatGPT. We collected each participant’s final program, then evaluated whether the programs satisfied their stated configurations (e.g., privacy budget).

We found that none of the novices were able to implement a DP mechanism that satisfied their stated configurations, whereas experts succeeded. 
Specifically, novices struggled to spot privacy issues in LLM-generated DP programs. Even when they sensed something might be wrong, they could not clearly articulate the problems and guide ChatGPT toward a fix. In contrast, we found experts verified and repaired the program iteratively. We then investigate how experts verify and iteratively repair the LLM-generated programs. Notably, we observed consistent patterns in the ways experts diagnose errors in code generated by large language models and prompt them to revise the program.

\noindent\textbf{RQ2: How can we scaffold the learning process to help students develop DP solutions with LLM support?}

We adopted a human-centered approach to iteratively design \name\ (Section~\ref{instructional-device-exploration}). We began with a graphical user interface to streamline the experts' workflow, guiding them to specify analysis goals and privacy configurations, followed by iteratively checking and correcting LLM-generated programs. However, users struggled to engage effectively due to a lack of background knowledge to make informed judgments. We then recognized that, instead of an interface, novices needed guidance on the underlying DP concepts and decision-making process. We switched to an example-based material (\name) that includes a worked example~\cite{renkl2014learning} and a set of erroneous examples~\cite{chen2019learning, tsovaltzi2010learning}. The worked example demonstrates an expert's step-by-step reasoning and decision-making process. The erroneous examples let novices practice evaluating choices at each key decision point in implementing DP.

We evaluated the effectiveness of \name\ with a between-subjects error identification user study (N=24) and an open-ended implementation study (N=5). In the error identification study, we evenly divided participants into two groups: one received \name\ and the other received a conventional instructional device that introduces basic DP concepts. We then compared their performance on identifying mistakes in two LLM-generated, flawed DP programs: one computes an average and the other computes z-scores (Section~\ref{verification-study}). To test whether they can transfer the strategies learned from \name\ to other data analysis use cases, we invited five participants from the \name\ group to return at least three days later for the open-ended implementation study. We gave them a sensitive dataset and asked them to (1) propose an analysis they would perform if they owned the data and wanted to share insights publicly, and (2) implement the analysis with differential privacy with LLM assistance (Section~\ref{open-ended-implementation}).

We found that participants who received \name\ identified an average of 66.7\% of the mistakes in LLM-generated DP programs. This was significantly more than the 15.0\% identified by participants who received the conventional instructional device. In the follow-up open-ended implementation study, participants proposed three analysis tasks: a z-test, linear regression, and multiple linear regression. During the implementation, they fixed an average of 63.3\% of the errors in the LLM-generated code, and two participants produced valid DP programs.

\noindent\textbf{Scope and Limitations:}
This paper presents exploratory work on instructional methods for novices implementing DP programs with the assistance of LLMs. Our work assumes a central differential privacy model~\cite{dwork2014algorithmic}, where raw data is sent to a trusted, centralized server. The practitioner develops a differentially private analysis that runs on the server.

\noindent\textbf{Contribution:}
Our primary contribution is introducing a pedagogical approach that brings worked and erroneous examples—well-established techniques from learning science—into the context of AI-assisted DP programming. 
Our exploration will help researchers understand how to teach novices to implement DP solutions effectively. To the best of our knowledge, this is the first study to systematically explore methods for teaching novices DP programming with the help of LLMs.

\section{Related Work}

\subsection{Tools for Implementing Differential Privacy}

Current DP tools target either technical users with DP knowledge or non-experts with minimal technical background. Tools for users with DP expertise include programming libraries~\cite{opendp2025, holohan2019diffprivlib, berghel2022tumult, pydp}, which provide reusable functions for implementing DP computations. These libraries facilitate DP programming by covering core operations like sensitivity computation~\cite{johnson2020chorus, wilson2019differentially}, privacy budget tracking~\cite{pipelinedp}, and standard mechanism implementations~\cite{holohan2019diffprivlib, autodp}. However, these tools are less usable for users who lack a solid understanding of DP~\cite{ngong2024evaluating, dibia2024sok, song2024inherently, panavas2024but}. Graphical interfaces like Private data Sharing Interface (PSI)~\cite{gaboardi2016psi} and DP Creator~\cite{dpcreator} support non-technical users through simplified parameter-setting processes and visual aids~\cite{dpwizard, Panavas2024VisualizationEducationDifferentialPrivacy, nanayakkara2022visualizing, nanayakkara2024measure}. However, research has shown that users without DP expertise often struggle to understand the reasons or implications of their choices~\cite{murtagh2018usable, sarathy2023don}. Our work targets the middle ground, supporting users with solid programming experience but limited exposure to DP in implementing differentially private data analyses.

\subsection{Educational Resources for Differential Privacy}

Researchers have worked on educational resources for DP practitioners through public tutorials~\cite{near_abuah_2021, ponomareva2023dp, desfontainesblog20181122, cowan2024hands} and university courses~\cite{cs860kamath, cs3110near, dsc291wang, cs2080Vadhan}. For instance, Programming Differential Privacy ~\cite{near_abuah_2021} is an open-source book that introduces DP concepts through code-based examples, guiding practitioners through implementation challenges step-by-step. These resources generally assume a conventional learning paradigm in which learners first acquire fundamental DP concepts before progressing to implementation. However, there is an increasing reliance on LLMs (e.g., ChatGPT) to generate programs, primarily using reference materials to interpret and correct model outputs~\cite{stackoverflow_2025_learning_to_code}. Our work aims to develop educational support tailored to this emerging paradigm.

\subsection{Using LLM on Domain-Specific Tasks}

Researchers have explored approaches to adapt LLMs for domain-specific tasks, including model adaptation, domain-specific frameworks, and human-centric scaffolding. Model adaptation involves fine-tuning LLMs on task-specific datasets~\cite{jeong2024fine, anisuzzaman2025fine, christophe2024med42}. Domain-specific frameworks integrate LLMs with external knowledge rather than modifying the model itself, often using retrieval-augmented generation (RAG)\cite{he2025using, barron2024domain, liang2025kag, zhang2025survey} or dictionary- or ontology-guided prompts\cite{zheng2024fine, potu2025ontology}. Human-centered scaffolding focuses on supporting users in leveraging LLMs for domain tasks, for example, through step-by-step scaffolds that help elicit correct domain-specific solutions~\cite{flores2025structured, tian2024theory} or adaptive prompt design~\cite{tang2025dynamic, lyu2023llm}. Building on this line of work, our paper focus on differential privacy specifically.

\section{A Needfinding Study: Can Developers Use LLMs to Implement Differential Privacy?} \label{need-finding}

We conducted a user study with six novices and three experts to examine whether they could use LLMs to implement DP solutions correctly and to identify the barriers they encountered.

\subsection{Recruitment \& Demographics}

We recruited participants through Slack channels and posted flyers at our university, as students are representative of developers when performing unfamiliar coding tasks ~\cite{tahaei2022recruiting, salman2015represent}. To avoid priming, we advertised the study as an LLM-assisted programming task rather than mentioning differential privacy. We screened participants through prior DP knowledge questions (Appendix~\ref{appendix:need-finding-prior-knowledge}) in the sign-up survey. We consider participants \textit{novices} if they (1) cannot name any DP techniques (e.g., the Laplace mechanism) and (2) have no DP implementation experience. We consider participants \textit{experts} if they (1) have heard of DP, (2) answered the privacy-budget interpretation question and the DP technique question correctly, and (3) have more than two years of experience in DP research.

We recruited participants on a rolling basis. In each round, we invited two novices and one expert and analyzed the collected data. The process continued until we no longer found new differences in implementation practices between the novices and the experts, indicating data saturation~\cite{saunders2018saturation}. We reached data saturation after three rounds, yielding six novices and three experts. Our novice group included four undergraduate students and two master's students, all majoring in data science or computer science. The expert group comprised one professor specializing in DP research and two PhD students who have hands-on experience applying DP in research projects. Table~\ref{tab:needfinding-participants} shows the participants' demographics and prior DP knowledge.

\begin{table}[h!]
\centering
\small
\caption{Summary of participants' demographics of our needfinding study.}
\begin{tabular}{|l| l l l l l| l l l l|}
\hline
ID & Group & Occupation & Dept. & Age & Gender & Aware DP & Small $\epsilon$ & DP Tech. & DP Res. Exp. \\
\hline
N1 & Novice & Undergrad & Data Sci. & 18--24 & Female & \checkmark & $\times$ & $\times$ & Never \\
N2 & Novice & Master's & Data Sci. & 18--24 & Male & $\times$ & $\times$ & $\times$ & Never \\
N3 & Novice & Undergrad & Data Sci. & 18--24 & Male & \checkmark & $\times$ & $\times$ & Never \\
N4 & Novice & Undergrad & Data Sci. & 18--24 & Male & \checkmark & \checkmark & $\times$ & Never \\
N5 & Novice & Master's & Com. Sci. & 18--24 & Male & \checkmark & $\times$ & $\times$ & Never \\
N6 & Novice & Undergrad & Data Sci. & 18--24 & Male & $\times$ & $\times$ & $\times$ & Never \\
\hline
E1 & Expert & PhD & Data Sci. & 18--24 & Male & \checkmark & \checkmark & \checkmark & 2--3 years \\
E2 & Expert & PhD & Data Sci. & 25--34 & Male & \checkmark & \checkmark & \checkmark & 2--3 years \\
E3 & Expert & Professor & Data Sci. & 35--44 & Male & \checkmark & \checkmark & \checkmark & >3 years \\
\hline
\end{tabular}
\label{tab:needfinding-participants}
\end{table}

\subsection{Study Protocol}

We conducted the study both in person and remotely via Zoom. We asked all participants to share their screen and record their audio in a Zoom meeting session to ensure consistent conditions. Participants began by completing a pre-survey to provide consent for participation and report their demographic information. Next, we asked them to set up their preferred Python programming environment and then presented the DP implementation task. We designed the task to simulate a ``data depositor'' setting~\cite{sarathy2023don}, where participants own the sensitive data and aim to share descriptive statistics publicly while preserving the data subjects' privacy. 

We provided a synthetic telemetry dataset and asked participants to implement a program that \textit{identifies product types with error rates larger than the average, where ``larger'' is defined as z-scores greater than 0} (see complete instructions in Appendix~\ref{appendix:task-description}), addressing any questions they had. We encouraged them to think aloud~\cite{van1994think} during the implementation to capture their decision-making process. To mirror real-world practice, we did not impose a time limit and encouraged participants to use ChatGPT. We considered the implementation complete when participants either felt confident in their solutions or chose not to continue. We collected the final program and the DP configuration they reported, including the privacy unit and total privacy budget. Finally, participants joined a post-interview to discuss their understanding of the program, the concepts they found confusing, and their reflections on the task. Both novices and experts followed the same protocol.

\subsection{Analysis Method}

We conducted thematic analysis~\cite{braun2006using} to identify challenges encountered by both novices and experts during their implementation of DP with LLMs. After each study session, two authors independently reviewed the recording and ChatGPT chat history, and open-coded individuals' interactions with LLMs, aiming to create as many codes as possible. Each code contained a participant's quote or a description of the observed behavior, along with the context. We then collaboratively combined codes into higher-level codes. Next, the two authors independently re-coded all sessions using the codebook and discussed disagreements. Finally, we merged similar codes into themes and sub-themes. Following prior work~\cite{mcdonald2019reliability}, we focused on generating emerging themes through iterative coding and discussion rather than measuring inter-rater reliability. Appendix~\ref{appendix:needfinding-codebook} presents our final codebook, which includes three themes, 10 sub-themes, and 34 codes.

We evaluated program correctness using three criteria: (1) executes without errors, (2) correctly implements the z-score algorithm, and (3) satisfies the participant-stated DP configurations. For example, a program claiming to achieve individual-level DP with a privacy budget of $1.0$ but actually consuming a budget of $2.0$ was deemed invalid.

\subsection{Ethical Considerations} \label{needfinding-research-ethics}
The study was approved by our Institutional Review Board (IRB). We used Zoom to record and transcribe the audio of all studies. After each study session, the authors checked and de-identified the transcription, then permanently deleted the audio recordings. All participants provided informed consent for the study procedures and the audio recording, and they acknowledged their right to withdraw at any time. Participants received a \$15 Amazon gift card and also gained educational benefits from engaging with the task. The compensation was approved by our IRB and aligns with the Belmont Report's Justice principle~\cite{united1978belmont}.

\subsection{Results}

All six novices submitted programs that failed to satisfy their stated DP configurations, whereas all three experts succeeded, even when starting from invalid LLM-generated code. From both groups, we identified three categories of barriers. Experts faced them less often and resolved them quickly, while novices struggled to overcome them.

\begin{table*}[h]
    \centering
    \small
    \caption{We identified three main barriers participants encountered when programming differentially private data analysis with LLMs. We present the complete codebook in Appendix~\ref{appendix:needfinding-codebook}.}
    \label{tab:barriers}
    \begin{tabular}{p{0.2\textwidth} p{0.18\textwidth} p{0.55\textwidth}}
        \toprule
        \textbf{Barrier Category} & \textbf{Issues} & \textbf{Examples} \\
        \midrule
        Struggles in setting DP configurations & Vague description for privacy requirements & \textit{(Prompt ChatGPT:) ``Propose a solution using differential privacy''} - N2 \\ 
        \midrule
        \multirow{4}{=}{Unable to verify LLM-generated solutions} & Require further clarification & \textit{``I should compute the maximum change (to set the sensitivity). But should it be a theoretical maximum change or the maximum change when one unit is added or removed from our current dataset?''} - N1 \\
        \cmidrule(lr){2-3}
        & Information overload & \textit{``There are too many things to figure out when implementing the program, so I could not find answers to all the questions in my mind.''} - N3 \\
        \cmidrule(lr){2-3}
        & \multirow{2}{=}{Confused by program outputs} & \textit{``Why are all the error rates 0s or 1s?''} - N1 \\ 
        &  & \textit{``The noisy z-score is pretty close to the actual z-score. Is that protecting user privacy?''} - N3 \\
        \midrule
        Difficulty in guiding LLMs to effective solutions & Lost in back-and-forth exploration & \textit{``Wait, why are the recapped steps adding noise to the counts. It was previously adding noise to the error rates.''} - N1 \\
        \bottomrule
    \end{tabular}
\end{table*}

\sssec{Struggles in setting DP configurations.} Most novices failed to decide on the privacy unit and the budget, two key elements of DP configuration that define the privacy guarantees~\cite{khavkin2025differential}. For example, N2 simply asked the model to \textit{``propose a solution using differential privacy.''} In such cases, LLMs arbitrarily assign a privacy budget and mix up privacy units, such as describing user-level DP but computing sensitivity at the event level. Experts also started without specifying clear DP configurations. However, they could quickly recognize the mismatch and redirect the model to the desired settings. For instance, E3 prompted, \textit{``I think what you got is good for event-level DP. But can we get user-level DP?''}

\sssec{Unable to verify LLM-generated solutions.} Novices relied on LLM-generated explanations to understand DP operations, but these were often unclear or overwhelming due to numerous new concepts. For instance, after reading about sensitivity, N1 asked, \textit{``Should it be a theoretical maximum change or the maximum change when one unit is added or removed from our current dataset?''} In contrast, N3 reflected in the post-interview, \textit{``ChatGPT explained too much beyond my task, and I felt tired halfway through.''} Experts also faced information overload. When E1 noted that clipping was needed, ChatGPT produced a long analysis of strategies and bounds, which E1 dismissed as ``too long.'' Despite this, E1 quickly zoomed in on the key decision: ``I want to make comparisons between clipping per user per category and per user vector.''

\sssec{Difficulty in guiding LLMs to effective solutions.} Novices often struggled with the iterative back-and-forth process, as LLM-generated solutions frequently introduced new DP concepts that required clarification and sometimes produced inconsistent programs when returning to the task. For example, N1 initially received a piece of code that injected Laplace noise into the error event counts and the total event counts. After two more interactions, the LLM suddenly started adding Laplace noise into error rates without any explanation, leaving N1 confused about where the noise should go. However, experts were less likely to ask clarification questions and instead directed the LLM to resolve program errors one at a time.

\section{Failures in LLM-Generated Differential Privacy Programs} \label{llm-failures}

To better understand the limitations of LLM code-generation for DP tasks, we tested GPT-4o to characterize the errors it produces. At the time of the study, GPT-4o was one of the state-of-the-art models.

\subsection{Method}

\sssec{Generate an initial implementation.} We tested the model on \textit{averages}, \textit{z-scores}, and \textit{one-way analysis of variance (ANOVA)} to cover different task complexity, each task repeated five times. 
Our prompt specified the analysis goal, data schema, and DP requirements, including the privacy unit (e.g., user-level), the neighboring dataset definition (e.g., add-or-remove-one user), and the total privacy budget (e.g., $\epsilon=1.0$). Appendix~\ref{appendix:llm-prompts} lists the complete prompts. We submitted the prompts through the ChatGPT web interface. To ensure independence between trials, we used separate chat sessions with memory disabled, creating 15 chat sessions in total.

\sssec{Evaluate program correctness \& code the mistakes.} We evaluated program correctness using three criteria: (1) executes without errors, (2) correctly implements the analysis algorithm, and (3) satisfies individual-level differential privacy with $\epsilon=1.0$. Two of the authors conducted thematic analysis~\cite{braun2006using} to identify patterns in the mistakes. The authors first reviewed all programs that did not satisfy DP line-by-line, open-coding each identified mistake. We then iteratively grouped similar codes into higher-level categories, re-coding each program after each refinement. The process continued until we reached a stable coding scheme, where no new codes emerged. We conducted two coding iterations to reach a consensus. The overall inter-coder agreement was 0.92, where we considered a program aligned only if the two raters identified the same set of mistakes. Eventually, we obtained five common mistake types. We will discuss each mistake type in the Section~\ref{llm-mistakes-result}.

\sssec{LLM self-check for potential mistakes.} To test whether LLMs could detect and correct errors, we prompted them with five previously identified mistake types, each illustrated with code examples. We submitted this self-check prompt to all chat sessions, regardless of the initial response's correctness level, then re-evaluated the updated programs. Next, we coded the mistakes in programs that did not satisfy DP using the five mistake types plus an `other' category for any new errors.

{
\newcommand{\cmark}{\cellcolor{correctbg}{\checkmark}}       %
\newcommand{\xmark}{\cellcolor{notdpbg}{$\times$}}           %
\newcommand{\warn}{\cellcolor{badutilitybg}{$\triangle$}}      %

\begin{table*}[h!]
\caption{LLMs are error-prone when implementing differentially private analysis. We tested GPT-4o on three tasks (5 trials each), and none of the programs satisfied DP. Even after explicitly listing common implementation mistakes and instructing the model to review and correct its code, the revised z-score and ANOVA programs still failed to satisfy DP.}

\small
\begin{tabular}{|l|cccc|cccc|cccc|}
\toprule
& \multicolumn{4}{c|}{\textbf{Average}} & \multicolumn{4}{c|}{\textbf{Z-score}} & \multicolumn{4}{c|}{\textbf{ANOVA}} \\
\cmidrule(lr){2-5} \cmidrule(lr){6-9} \cmidrule(lr){10-13}
\textbf{Trial} 
& \multicolumn{2}{c}{Init. Impl.} & \multicolumn{2}{c|}{Post Self-check}
& \multicolumn{2}{c}{Init. Impl.} & \multicolumn{2}{c|}{Post Self-check}
& \multicolumn{2}{c}{Init. Impl.} & \multicolumn{2}{c|}{Post Self-check} \\
\cmidrule(lr){2-3} \cmidrule(lr){4-5}
\cmidrule(lr){6-7} \cmidrule(lr){8-9}
\cmidrule(lr){10-11} \cmidrule(lr){12-13}
& Corr. & Mistake & Corr. & Mistake 
& Corr. & Mistake & Corr. & Mistake 
& Corr. & Mistake & Corr. & Mistake \\
\midrule
\#1 & \xmark & 1, 2 & \cmark & -- 
     & \xmark & 5 & \xmark & 5 
     & \xmark & 5 & \xmark & 5 \\
\#2 & \xmark & 3 & \cmark & -- 
     & \xmark & 1, 5 & \xmark & 1, 5 
     & \xmark & 2 & \xmark & 5 \\
\#3 & \xmark & 3 & \cmark & -- 
     & \xmark & 1, 5 & \xmark & 1, 5 
     & \xmark & 5 & \xmark & 5 \\
\#4 & \xmark & 1, 4 & \cmark & -- 
     & \xmark & 5 & \xmark & 5 
     & \xmark & 1, 5 & \xmark & 1, 5 \\
\#5 & \xmark & 2, 4 & \cmark & -- 
     & \xmark & 5 & \xmark & 5 
     & \xmark & 3 & \xmark & 2 \\
\bottomrule
\end{tabular}

\vspace{1mm}
\begin{tablenotes}
\item 
\begin{tabular}{lll}
1. Misused Sensitivity & 2. Unprivatized data-dependent hyper-parameters & 3. Only partially privatized \\
4. Overly noisy result & 5. Exceeded privacy budget & \\
\end{tabular}
\end{tablenotes}

\label{tab:llm-mistakes}
\end{table*}
}

\subsection{Result}~\label{llm-mistakes-result}

\sssec{All initial implementations failed to satisfy DP.} Across all programs, we observed five mistake types, each recurring in multiple implementations (Table~\ref{tab:llm-mistakes}). These include:

\begin{enumerate}[noitemsep,topsep=3pt,leftmargin=*]
    \item \textbf{Misused sensitivity.} The sensitivity is not the maximum change in the result when one privacy unit is added or removed. For example, Average\#4 sets \texttt{sensitivity = 1}, though the average could shift by up to $7$ when one privacy unit (= an individual) is added or removed.
    
    \item \textbf{Unprivatized data-dependent hyperparameter.} Setting hyperparameters such as sensitivity by querying the data without noise protection violates differential privacy. For instance, Average\#5 sets \texttt{sensitivity = 7 / len(clipped\_counts)} while \texttt{clipped\_counts} were obtained by transforming the sensitive dataset.
    
    \item \textbf{Only partially privatized.} Differential privacy guarantee fails when noise is not protecting every paths in the computation graph. For example, Average\#2 computes \texttt{private\_average = private\_sum / true\_count} while the count in the denominator is not privatized.
    
    \item \textbf{Overly noisy result.} Adding noise to queries with a large signal-to-sensitivity ratio. Though this mistake itself does not disqualify a program from satisfying DP, it may cause the program to provide poor utility. For example, Average\#4 computes the noisy mean by adding noise to the true mean (\texttt{noisy\_mean = true\_mean + np.random.laplace(loc=0, scale=laplace\_scale)}). However, this causes the results to be noisier than adding noise to the numerator and the denominator separately.
    
    \item \textbf{Exceeded privacy budget.} Privacy budgets accumulate across differentially private releases. Failing to allocate budgets properly can result in exceeding the total privacy budget constraint. For example, Z-score\#3 consumes $\epsilon = 1.0$ to compute the error rate for a single product type, which leads to a total consumption of $\epsilon = 7.0 > 1.0$ when extended to all seven product types.
\end{enumerate}

\sssec{Self-check only fixes five of the programs (33.3\%).} The remaining 10 programs still failed to satisfy DP. The model successfully fixed only the simpler cases, namely the five programs computing averages. Among the 10 programs that did not satisfy DP, eight retained their original mistakes and two introduced new errors. For example, ANOVA\#2 originally had unprivatized data-dependent hyperparameters, but after self-check, the new version exceeded the privacy budget. Notably, no new mistake types beyond the five originally identified appeared in the new programs.

\section{Iterative Development of \name} \label{instructional-device-exploration}

To lower the barrier to DP implementation, we adopted a human-centered approach to iteratively design instructional materials that scaffold students’ learning to implement DP with LLMs.

\subsection{First Attempt: Interactive LLM-Driven DP Programming Interface} \label{verification-checklist}

\begin{figure}[h]
    \centering
    \includegraphics[width=0.45\linewidth]{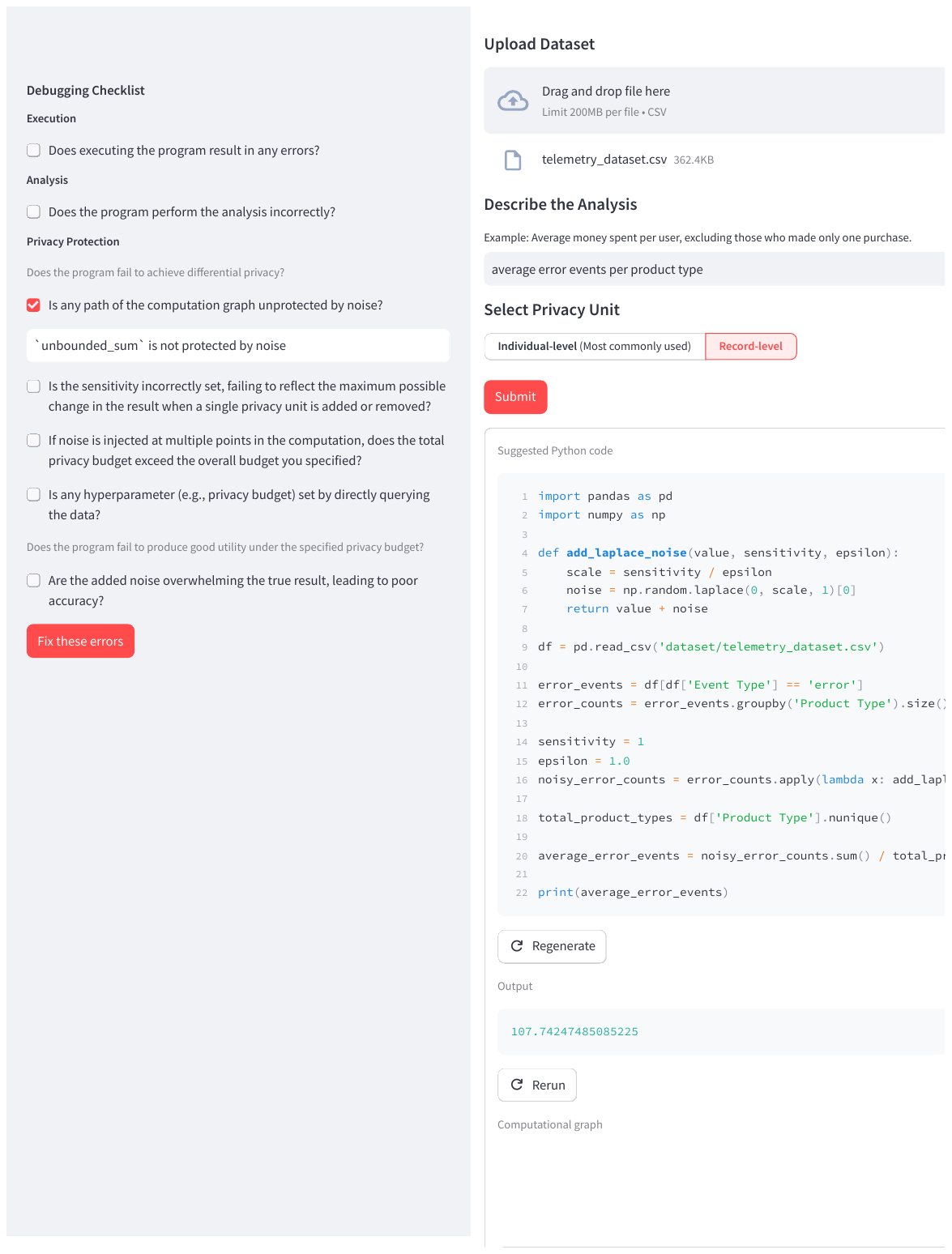}
    \caption{We initially designed an interactive LLM-driven DP programming interface, which includes a debugging checklist that directs users to key parts where mistakes often arise. The checklist covers five common mistakes that invalidate DP compliance.}
    \label{fig:gui-checklist}
\end{figure}

We began with a graphical user interface (GUI) approach, a well-established workflow scaffolding method demonstrated in DP Wizard~\cite{dpwizard} and DP Creator~\cite{dpcreator}. We designed our initial prototype as a GUI that streamlines the expert workflow of developing DP programs with LLMs, from specifying requirements to iteratively verifying and correcting the LLM-generated code.

The GUI first asks users to enter the requirements through a form, including the analysis goal (e.g., average spending per user), dataset upload, total privacy budget (e.g., $\epsilon = 1.0$), and privacy unit (e.g., individual). A GPT-powered backbone then generates the program along with a short explanation and a flow graph, both designed to help users understand how the program works (Appendix~\ref{appendix:GUI-screenshots}). Next, a debugging checklist appears, directing users to key parts where mistakes often arise. The checklist lists five key mistake types identified in Section~\ref{llm-mistakes-result} and basic checks for execution and analysis correctness. Users flag unmet criteria, describe the observed errors, and request repairs (Figure~\ref{fig:gui-checklist}). The cycle repeats until the program satisfies all checks. We implemented the GUI as a Streamlit web application and built the GPT-powered backend on OpenAI’s GPT-4o real-time API.

We tested the GUI with four pilot users: one undergraduate student in data science, one graduate student in computer science, and two data science professors. One professor had rich implementation experience in DP, helping us ensure the GUI supports DP practices correctly. The other users had limited or no DP implementation experience, which allowed us to observe how novices interact with the system.
We observed that the workflow forces users to specify concrete privacy requirements, preventing novices from leaving units and budgets to the LLM. However, novices lack the DP knowledge needed to make informed judgments about the debugging checklist.

\subsection{Second Attempt: Worked Example}
Through the first attempt, we realized the challenge lies in supporting novices’ learning, not the interface itself. Novices are often unaware of the decision process and require substantial effort to understand the jargon. For example, deciding the privacy unit requires understanding the concept of ``neighboring datasets'', which represents datasets differing by the addition or removal of one unit---an abstract idea that beginners rarely find intuitive.

We then stepped back to examine the decision space of implementing differentially private data analysis across descriptive statistics, hypothesis testing, and regression methods. Table~\ref{tab:data-analysis-methods} lists the analyses we explored. Surprisingly, despite the variety of analyses, implementing differential privacy with LLMs follows a similar sequence of decisions. As an illustrative example, we outline the decisions for implementing differentially private z-scores, the task used in our prior needfinding study (Section~\ref{need-finding}):

\begin{table}[h]
\centering
\small
\caption{We analyzed statistical data analysis methods covering descriptive statistics, hypothesis testing, and correlation/regression. Implementing these analyses with differential privacy follows a similar decision-making pattern.}
\begin{tabular}{|l|l|}
\hline
\textbf{Category} & \textbf{Analysis Methods} \\
\hline
Descriptive statistics & mean, weighted mean, standard deviation, proportions, z-scores~\cite{abdi2007z} \\
Hypothesis testing & t-test~\cite{kim2015t}, analysis of variance (ANOVA)~\cite{st1989analysis}, chi-square test~\cite{franke2012chi}, Mann--Whitney U test~\cite{mcknight2010mann} \\
Correlation \& regression & Pearson correlation~\cite{benesty2009pearson}, linear regression~\cite{montgomery2021introduction}, multiple linear regression~\cite{tranmer2008multiple} \\
\hline
\end{tabular}
\label{tab:data-analysis-methods}
\end{table}

\begin{enumerate}[noitemsep,leftmargin=*]
    \item \textbf{What DP protection should we provide?} We aim to protect each individual user's privacy, keeping the total privacy budget within $\epsilon_{total} = 1.0$.
    \item \textbf{Is the program performing the desired analysis?} The program computes error rates per product type and then derives the z-scores.
    \item \textbf{Where is noise added?} The program adds noise to both the total and error event counts for each product type. Great, the resulting z-scores would also satisfy DP by the \emph{post-processing property}.
    \item \textbf{What is the sensitivity of the queries?} The event counts are unbounded by nature. We should \emph{clip} them to enforce a bound. Let's clip the number of events per user to 5.
    \item \textbf{How is the privacy budget allocated?} The total and error events queries overlap on end users. With seven product types and two queries each, each query consumes a privacy budget of $\epsilon_{\text{total}}/14$.
    \item \textbf{Does the result make sense?} Noise can make the total and error event counts negative. Let's add a small number to avoid ``divide by near zero'' and clip error rates to keep them within 0--1.
    \item \textbf{Is the result too noisy?} The \emph{signal-to-sensitivity} ratio exceeds $1$ if each product type has at least $5$ error events, which is reasonable for a large dataset. The result should not be too noisy.
\end{enumerate}

This motivates us to adopt a learn-by-example approach~\cite{pirolli1985role, van2010example}.
We employed \textbf{worked examples}, a well-established learning science principle in which learners learn from step-by-step demonstrations of the process required to complete a task~\cite{renkl2014learning, atkinson2000learning, muldner2022review}.
In our case, the worked example illustrates the key decision points in implementing DP with LLMs and the choices available at each step.
Prior work found that observing structured demonstrations helps learners avoid ineffective strategies, such as aimless trial-and-error or superficial pattern-matching, and focus on the core principles needed to solve the task~\cite{mitWorkedExamples}.

\begin{figure*}[t]
    \centering
    \includegraphics[width=1\linewidth]{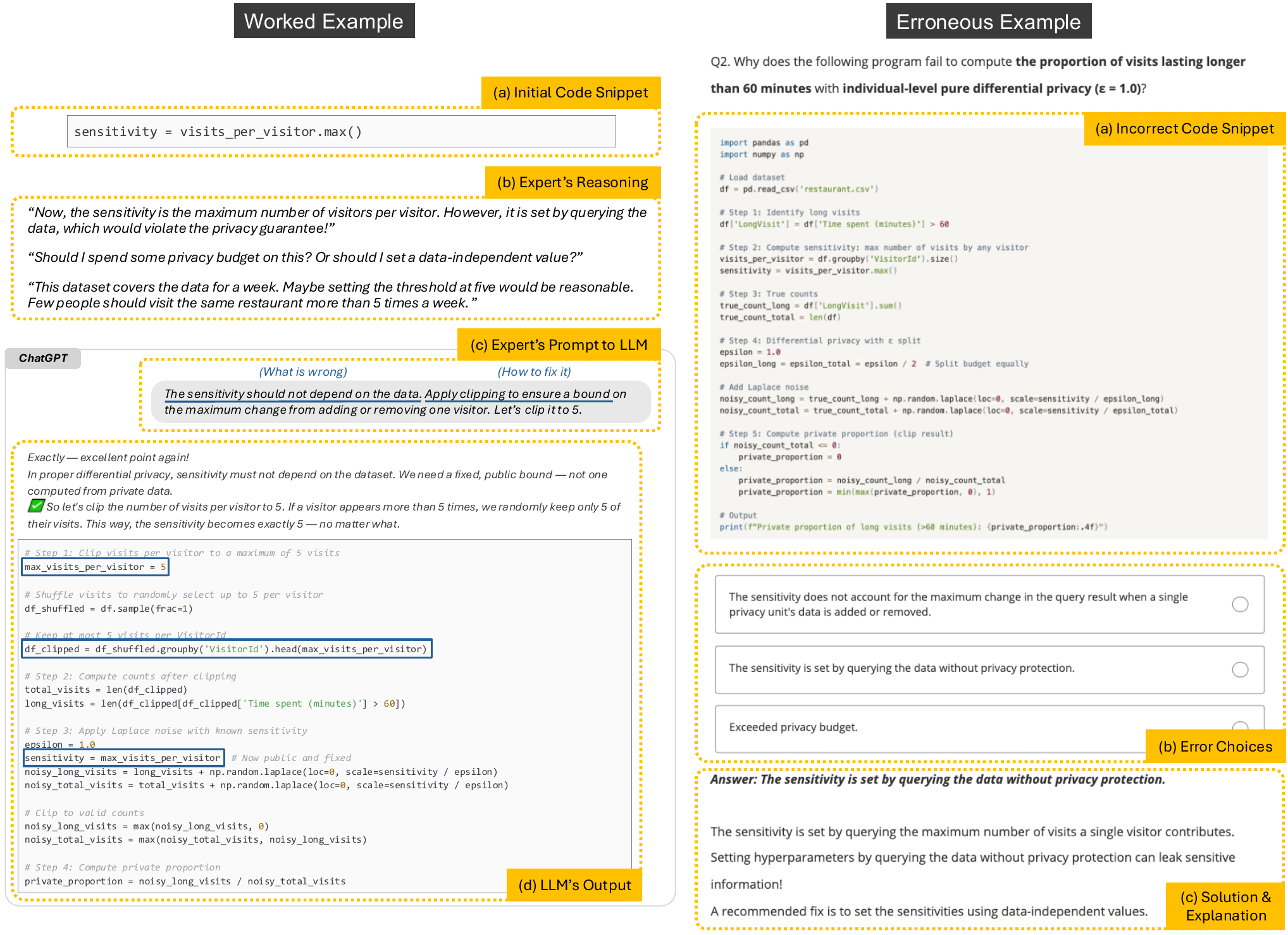}
    \caption{Inspired by learning science principles, \name\ includes a \textbf{Worked Example} (left), which demonstrates an expert’s step-by-step problem-solving approach, and \textbf{Erroneous Examples} (right), which let learners practice evaluating choices at each key decision point.}
    \label{fig:worked_and_erroneous_examples}
\end{figure*}

We designed our worked example as a tutorial video that step-by-step demonstrates how an expert develops a DP program with an LLM. We built the example around a differentially private ratio scenario--given a restaurant visit dataset, compute the proportion of customer visits lasting longer than 60 minutes while satisfying individual-level pure differential privacy with a total privacy budget $\epsilon_{total}=1.0$. The tutorial video presents the expert's decision process from a first-person perspective. It begins with the expert prompting to generate an initial DP program. It then steps through potential errors, highlighting the code under inspection and articulating the expert’s reasoning for each check as they would when mentoring a novice. When an error is present, the expert demonstrates the prompt used to fix it and shows the corrected code (Figure~\ref{fig:worked_and_erroneous_examples} left).

We tested the worked example with the same pilot users as the previous GUI. This allowed us to compare the two approaches while controlling for individual differences. Our observation confirmed that worked examples helped learners follow the structured decision-making process. However, learners passively followed the expert's decisions without actively reasoning about the choices at each decision-making point.

\subsection{Final Design: Worked Example \& Erroneous Examples (\name)} \label{worked-example-erroneous-examples}

In the second attempt, the worked example demonstrates the decision-making points but offers limited opportunities for learners to reason about the choices at each step. We filled this gap by incorporating \textbf{erroneous examples} that provide targeted practice to evaluate and select the correct options. Previous studies have shown that engaging with erroneous examples can deepen learners’ conceptual understanding, especially when they are encouraged to actively diagnose and correct mistakes~\cite{wang2022learning, tsovaltzi2010learning, richey2019more}.

We designed each erroneous example as a flawed LLM-generated DP program requiring students to identify the underlying mistake. Each program contains a single error to give learners simple, focused practice. To minimize overhead from understanding a new data analysis scenario, we built the programs around the same differentially private ratio scenario as in the worked example. We generated all programs by GPT-4o to ensure the examples reflect realistic mistakes that learners may encounter when using LLMs for DP programming. We created five examples targeting common LLM mistakes, focusing on (1) sensitivity definition, (2) data-dependent hyperparameters, (3) partially privatized computation, (4) noise scale, and (5) privacy budget composition.

We implemented the erroneous examples as multiple-choice questions using Qualtrics. Each question contains three choices, each corresponding to a type of mistake.
Through the survey, students see the correct solution and a concise explanation after each question (Figure~\ref{fig:worked_and_erroneous_examples} right), as prior research demonstrates the effectiveness of timely over delayed feedback~\cite{kulik1988timing}.
At the end of the survey, students receive a full review of all questions to support reflection and reinforce learning~\cite{kim2020partial}.

We tested \name\ on the same group of pilot users and confirmed that the erroneous examples effectively prompted learners to actively think about each decision choice. The time overhead was minimal. Participants progressed at their own pace and completed all erroneous examples in approximately 10 minutes.

\section{Evaluation} \label{evaluation}

Our evaluation seeks to answer the following two questions:
\begin{itemize}[noitemsep,topsep=3pt,leftmargin=*]
    \item Do DP novices trained with \name\ perform better at identifying implementation issues in DP data analysis programs than those trained with a conventional instructional method?
    \item Can DP novices generalize the errors demonstrated in \name\ to other data analysis use cases?
\end{itemize}
We conducted two user studies, an error identification study (N=24) (Section~\ref{verification-study}) and an open-ended implementation study (N=5) (Section~\ref{open-ended-implementation}). Both studies were conducted primarily at our university and approved by our IRB.

\subsection{Error Identification Study} \label{verification-study}

The study evaluates whether \name\ helps DP novices identify issues in LLM-generated DP programs better than conventional training. Following a between-subjects study design, we divided participants evenly into two groups. The \name\ group received \name\ training. The baseline group followed a conventional self-learning path, using a handout as a stand-in for tutorial websites and blogs, a tutorial video, and a practice quiz.

\begin{figure}[h]
    \centering
    \includegraphics[width=0.7\linewidth]{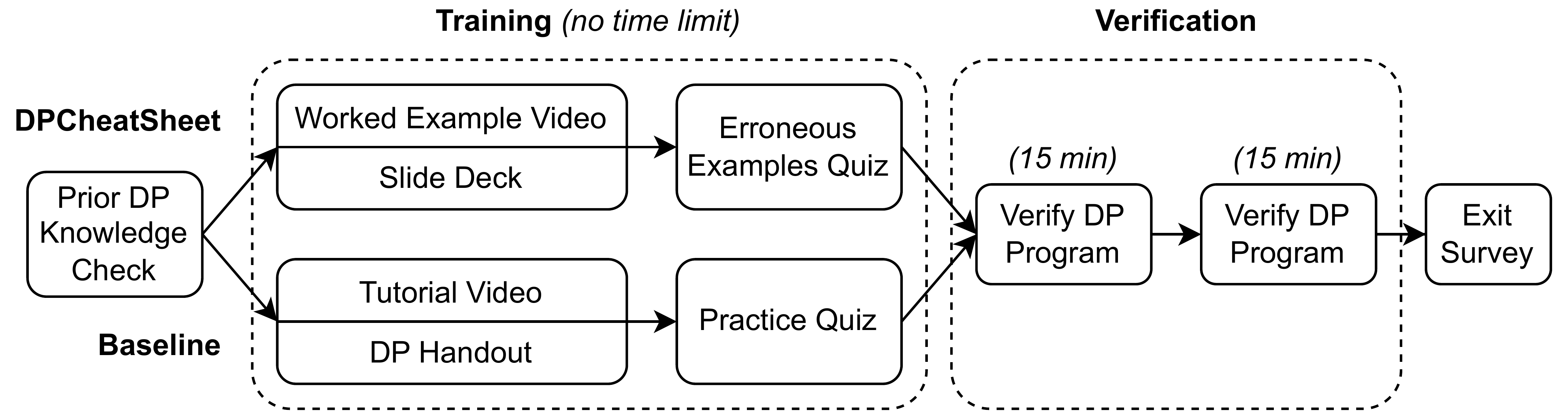}
    \caption{Error identification study procedure overview.}
    \label{fig:verification-study-procedure}
\end{figure}

\sssec{Study procedure.} Each session began with participants providing consent for participation. They then completed a prior DP knowledge questionnaire adapted from Ngong~\cite{ngong2024evaluating} (Appendix~\ref{appendix: verification-priordp}) and proceeded to the training phase.

Participants in the \name\ group watched the worked example video with the slide deck they could review at their own pace. They then completed a five-question quiz with five multiple-choice questions, each presenting a flawed DP program and asking participants to identify the error. Participants in the baseline group, instead, watched a tutorial video on applying Laplace noise to achieve DP\footnote{https://www.youtube.com/watch?v=mO0NmkVGapM}, accompanied by a handout adapted from~\cite{ngong2024evaluating}. They then completed a five-question quiz selected from the midterm exam of a quarter-long DP course. In both groups, participants received immediate feedback with explanations after each question, access to the tutorial materials, and no time limit. Training took an average of 24.3 minutes (SD = 5.5) in the \name\ group and 12.4 minutes (SD = 2.5) in the baseline group.

After training, participants proceeded to the error identification tasks, which involved two flawed ChatGPT-generated DP programs: one computes a differentially private average, and another computes differentially private z-scores (Appendix~\ref{appendix:verification-programs}). The average program misused sensitivity, privatized only part of the computation, and exceeded the privacy budget. The z-score program relied on unprivatized data-dependent hyperparameters and produced overly noisy results. We asked participants to identify all errors preventing the programs from satisfying user-level DP under the add-or-remove-one definition ($\epsilon = 1.0$) and briefly explain the mistakes. Participants did not use LLMs during the task. We counterbalanced the order of the tasks and session format (in-person vs. online). We imposed a 15-minute time limit per task and recorded participants' responses and the time spent on each task.

At the end of the study, participants completed an exit survey including the NASA Task Load Index (NASA-TLX)~\cite{hart1988development} to report perceived workload, and open-ended questions to explain whether and why the materials were helpful. Finally, we debriefed participants, explaining all underlying errors in the two tasks. Each session lasted approximately one hour. Figure ~\ref{fig:verification-study-procedure} illustrates the study procedure. 

\sssec{Participant recruitment and demographics.} We recruited 24 students by advertising on campus-wide Discord servers, Slack channels, and physical flyers posted around campus. We pre-screened participants to ensure they (1) were at least 18 years old, (2) had at least one year of programming experience, and (3) had prior experience with LLMs. We intentionally included participants with varying levels of prior experience in DP. Each participant received a \$15 Amazon gift card, with an additional \$15 awarded to the top three performers in each group. Participants also gained educational benefits from engaging with the tasks. The compensation was approved by our IRB and aligns with the Belmont Report's Justice principle~\cite{united1978belmont}.

Our final group included 12 undergraduate students, 7 master's students, and 5 PhD students. Seventeen participants reported not knowing what DP means, 6 had some experience with DP (e.g., an introductory class or research papers), and 1 had research experience with DP. Eight participants (33.3\%) identified as female, and the rest as male. Fourteen (58.3\%) participants were aged 18--24, and 10 (41.7\%) were 25--30. Twelve participants (50\%) majored in Computer Science, 6 (25\%) in Data Science, 3 (12.5\%) in Electrical and Computer Engineering, 2 (8.3\%) in Cognitive Science, and 1 (4.2\%) in Physics.

\sssec{Analysis method.} We first coded the 48 responses (2 questions × 24 participants) based on the five common mistakes observed in LLM-generated programs (Section ~\ref{llm-mistakes-result}). For example, in response to the average program, one participant wrote: \textit{(1) The \texttt{count} variable should be added with noise as well. We can split the epsilon for \texttt{total} and \texttt{count} and then add noise to both of them. (2) One user could correspond to multiple events, so sensitivity should not be set to one. (3) The noisy average should be cropped to $[0, 1]$ to ensure validity.} We coded the first two items as correctly identified errors: \textit{Privatized only part of the computation} and \textit{Misused sensitivity}, and the third item as an incorrectly flagged mistake: \textit{Incorrect post-processing for data validity}.

We calculated each participant's precision and recall. \textbf{Precision} is defined as the proportion of correctly identified mistakes out of all the issues the participant pointed out. \textbf{Recall} is computed as the proportion of correctly identified mistakes out of the five actual mistakes present in the two programs.

We then performed two multiple linear regressions with precision and recall as the dependent variables. Independent variables included instructional device received (\name\ vs. baseline), program order (z-score first vs. average first), training format (online vs. in-person), and participants' prior knowledge scores (ranging from 0-4). We dummy-coded the binary categorical variables (instructional device, program order, format) into 0s and 1s.

We also analyzed the difficulty of each differential privacy concept by counting how many participants correctly identified each of the five mistakes. Finally, we conducted Mann–Whitney U tests to compare \name\ and baseline participants across the NASA TLX scales.

\subsubsection{Quantitative Results}

\begin{table}[h]
\centering
\caption{Coefficients and p-values for the multiple linear regression predicting \textbf{\textit{precision}}. A positive coefficient with a small p-value indicates that the factor increases the precision. Participants who received \name\ achieved significantly higher precision ($p < 0.001$). More of their identified issues were actual mistakes.}
\begin{tabular}{lcc}
\toprule
\textbf{Variable} & \textbf{Coefficient} & \textbf{P-value} \\
\midrule
\textbf{Instr. Device}: \name\ (vs. Baseline)          & \textbf{0.659} & \textbf{0.000}\textsuperscript{***} \\
\textbf{Program Order}: Z-score first (vs. Avg first) & -0.073 & 0.439 \\
\textbf{Format}: In-person (vs. Online)               & 0.059 & 0.537 \\
\textbf{Prior Knowledge Score}                        & 0.005 & 0.908 \\
\textbf{Intercept}                                    & 0.172 & 0.123 \\
\bottomrule
\multicolumn{3}{l}{
  \parbox{0.6\linewidth}{
    $R^2 = 0.732$, Adjusted $R^2 = 0.676$\hfill ***$p < 0.001$
  }
}
\end{tabular}
\label{tab:precision-regression}
\end{table}

\begin{table}[h]
\centering
\caption{Coefficients and p-values for the multiple linear regression predicting \textbf{\textit{recall}}. A positive coefficient with a small p-value indicates that the factor increases the recall. Participants who received \name\ achieved significantly higher recall ($p < 0.001$). They identified a greater proportion of all actual mistakes.}
\begin{tabular}{lcc}
\toprule
\textbf{Variable} & \textbf{Coefficient} & \textbf{P-value} \\
\midrule
\textbf{Instr. Device}: \name\ (vs. Baseline)          & \textbf{0.514} & \textbf{0.000}\textsuperscript{***} \\
\textbf{Program Order}: Z-score first (vs. Avg first) & 0.051 & 0.554 \\
\textbf{Format}: In-person (vs. Online)               & 0.014 & 0.873 \\
\textbf{Prior Knowledge Score}                        & 0.011 & 0.769 \\
\textbf{Intercept}                                    & 0.102 & 0.306 \\
\bottomrule
\multicolumn{3}{l}{
  \parbox{0.6\linewidth}{
    $R^2 = 0.665$, Adjusted $R^2 = 0.595$\hfill ***$p < 0.001$
  }
}
\end{tabular}
\label{tab:recall-regression}
\end{table}

Table~\ref{tab:precision-regression} shows the regression results for precision. Among the four independent variables, only \name\ intervention had a significantly positive effect on precision scores ($\beta = 0.659$, $p < 0.001$). Since the other variables did not show significant effects, we conducted a follow-up comparison of precision scores between the two instructional conditions (\name\ vs. baseline). As precisions were not normally distributed (Shapiro--Wilk $p < 0.01$), we applied a Mann--Whitney U test. The result showed a significant difference in precision scores between the two groups ($p = 0.000$), suggesting that \name\ led to higher precision scores than the baseline method.

Table~\ref{tab:recall-regression} presents the regression results for recall. Again, only \name\ intervention had a significant effect ($\beta = 0.514$, $p < 0.001$) on recall, indicating that \name\ helped students identify more actual mistakes compared to the baseline. As recall scores were non-normally distributed (Shapiro--Wilk $p < 0.01$), a Mann–Whitney U test shows a significant group difference ($p = 0.000$).

\begin{figure}[h]
    \centering
    \includegraphics[width=0.65\linewidth]{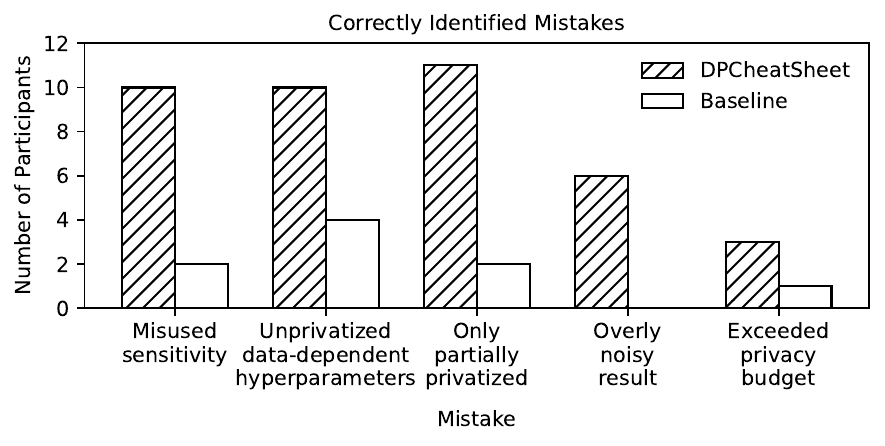}
    \caption{Number of participants in each group who correctly identified each mistake. Participants in the \name\ group consistently outperformed the baseline group, with more than half identifying \textit{misused sensitivity}, \textit{unprivatized data-dependent hyperparameters}, \textit{only partially privatized}, and \textit{overly noisy result}.}
    \label{fig:true-positives}
\end{figure}

Among the five underlying issues, participants receiving \name\ performed best at identifying \textit{partially privatized logic}, with 11 participants (92\%) catching it. \textit{Misuse of sensitivity} and \textit{unprivatized data-dependent hyperparameters} were each identified by 10 participants (83\%). Six participants (50\%) correctly spotted \textit{Overly noisy results}, and 3 participants (25\%) caught\textit{exceeded privacy budgets}. In comparison, fewer participants in the baseline group identified each of the five mistakes. The most commonly identified error in the baseline group was \textit{unprivatized data-dependent hyperparameters}, with 4 participants (33\%) recognizing it. None of the baseline participants noticed \textit{overly noisy results}.

\begin{table*}[h!]
\centering
\caption{Participants’ responses to the NASA TLX survey, reported as median (mean $\pm$ standard deviation) on a 7-point Likert scale. Arrows indicate the desired direction for each measure (\(\uparrow\) denotes that a higher value is better, and \(\downarrow\) denotes that a lower value is better). For each measure, we highlighted the value with the better median in bold.}
\small
\begin{tabular}{lcccccc}
\toprule
\makecell{\textbf{Instructional}\\\textbf{Device}} & \textbf{Mental Dem.~\(\downarrow\)} & 
\textbf{Physical Dem.~\(\downarrow\)} &
\textbf{Temporal Dem.~\(\downarrow\)} & 
\textbf{Performance~\(\uparrow\)} & 
\textbf{Effort~\(\downarrow\)} & 
\textbf{Frustration~\(\downarrow\)} \\
\midrule
\name        & 5.5 (5.6 $\pm$ 0.9) & \textbf{2.0 (2.4 $\pm$ 1.6)} & 3.5 (3.3 $\pm$ 1.2) & \textbf{4.0 (3.9 $\pm$ 1.4)} & 5.0 (5.3 $\pm$ 0.7) & \textbf{2.5 (3.0 $\pm$ 1.5)} \\
Baseline     & \textbf{5.0 (5.5 $\pm$ 0.7)} & 2.5 (2.6 $\pm$ 1.4) & \textbf{3.0 (3.0 $\pm$ 1.8)} & 2.5 (3.0 $\pm$ 1.5) & 5.0 (4.9 $\pm$ 1.2) & 5.0 (4.2 $\pm$ 1.5) \\
\bottomrule \\[-4pt]
\end{tabular}
\label{tab:nasa_tlx}
\end{table*}

We present the results of the NASA TLX survey in Table~\ref{tab:nasa_tlx}. Compared to the baseline, participants using \name\ reported lower physical demand and frustration, and higher perceived performance based on median scores. Mann–Whitney U tests on all six scales revealed no statistically significant differences between the groups.

\subsubsection{Qualitative Results}

To understand how \name\ supported participants in achieving better performance, we analyzed their feedback in the exit survey which led to the following key findings.

\sssec{The worked example provides a clear workflow with concrete steps to follow.} Several participants noted that the tutorial video outlined a structured process that helped them navigate the verification task. Participant D11 said it ``\textit{clearly conveyed the general steps that I should follow in evaluating differential privacy, pointed out the parts of the code that I should check for errors.}'' Similarly, participant D12 shared that ``\textit{the tutorial video generally go[es] through the whole workflow which gives me a mindset to solve the problem.}''

\sssec{The erroneous examples encourage sense-making.} Participants highlighted how these examples helped them engage more deeply with the material by clarifying any misunderstandings. For instance, participant D10 noted that the examples ``\textit{show which part I understand wrong,}'' helping them bridge their knowledge gaps. Participant D11 shared, ``\textit{The practice quiz mainly helped me in reinforcing the concepts from the tutorial video; additionally, it helped me with understanding a concept that I had glossed over in the tutorial video.}''

\subsubsection{Why many participants miss the \textit{exceeded privacy budget} error?} In the debrief, several participants in the \name\ group noted that they assumed the seven differentially private releases, one for each product type, could be treated as parallel since they involve disjoint product categories. However, this interpretation overlooks the fact that users who contributed data across all product types may overlap. This distinction between disjoint data values and disjoint users can be subtle and may be particularly challenging for students to grasp.

\subsection{Open-Ended Implementation Study} \label{open-ended-implementation}

We invited five participants from the \name\ group in our previous error identification study (Section~\ref{verification-study}) to return at least three days after their initial session. In the follow-up study, they implemented a self-chosen data analysis task with DP.
Our goal was to examine whether \name\ can help novices internalize expert workflows for DP implementation and apply them to other data analysis tasks.

\begin{figure}[h]
    \centering
    \includegraphics[width=0.6\linewidth]{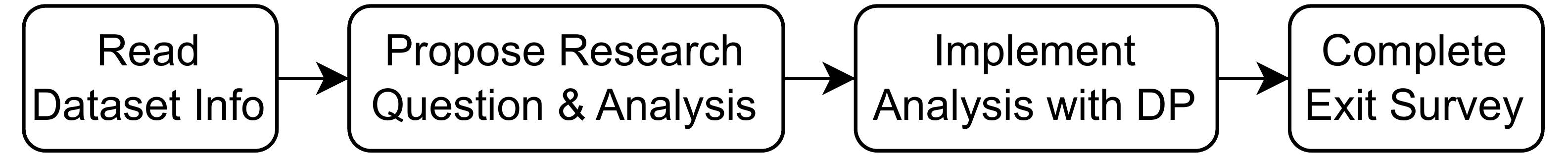}
    \caption{Open-ended implementation study procedure overview.}
    \label{fig:field-study-procedure}
\end{figure}

\sssec{Study procedure.} We conducted all sessions in person and recorded participants’ screens via Zoom for further analysis. Each session lasted approximately one hour. The study contains two main parts: (1) proposing an analysis and (2) implementing the proposed analysis with differential privacy (Figure~\ref{fig:field-study-procedure}).

We began by presenting participants with a Student Social Media \& Relationships dataset~\cite{kaggle_dataset}. The dataset contains anonymized records of students’ social media behaviors and associated aspects of their personal lives, such as their average sleep hours per night and self-rated mental health scores. We then prompted them with the following scenario: ``\textit{Imagine you are a researcher. You collected survey responses from students about their social media usage, mental health, and relationships. You would now like to share some insights from this data publicly, such as in a blog post or report.}''

We first asked participants to decide what insights they would like to share from the dataset. We instructed them to write down a research question (e.g., do younger students spend more time on social media?) and specify a corresponding analysis method (e.g., compare the average daily usage hours across age groups using a t-test). We then asked participants to implement the analysis, ensuring it satisfied individual-level differential privacy with a total budget of $\epsilon = 1.0$ to protect the students’ privacy. We encouraged them to use ChatGPT to write and refine their code as demonstrated in the worked example. We also provided printed materials, including the worked example slide deck, erroneous examples, and the tutorial video, which they could freely refer to as needed. The task ended when participants indicated they were satisfied with their solution. There was no time limit, and we recorded the duration of each implementation.

At the end, participants completed an exit survey. It included the NASA TLX~\cite{hart1988development} to assess perceived workload and open-ended questions about challenges they faced and any potential mistakes they suspected in their solutions.

\sssec{Recruitment.} We emailed the 7 participants in the \name\ group who had agreed to be contacted for follow-up, inviting them to a one-hour, in-person study; 5 accepted and participated. The sample size is similar to end-to-end programming user studies in prior research, which typically recruit a small number of participants due to the time-consuming and cognitively demanding nature of the tasks~\cite{li2017privacystreams, jin2022peekaboo}.

Among the five participants, three were undergraduate students and two were master's students. Before our previous session, four participants (80\%) did not know what DP means, and one participant (20\%) had some experience with DP. Four identified as female, and the others as male. Four participants were aged 18--24 and one was 25--30. Three participants (60\%) majored in Computer Science and two (40\%) in Data Science. Each participant received a \$20 Amazon gift card as compensation and also gained educational benefits from completing the task. The compensation was approved by our IRB and aligns with the Belmont Report's Justice principle~\cite{united1978belmont}.

\begin{table*}[h]
\centering
\caption{Participants in the open-ended implementation study proposed five different research questions to the same student dataset. They selected analysis methods ranging from Z-tests to linear regressions.}
\label{tab:rq-and-analysis}
\small %
\begin{tabularx}{\textwidth}{@{}l X l@{}}
\toprule
\textbf{ID} & \textbf{Research Question} & \textbf{Analysis Method} \\
\midrule
P1  & Does different most-used platforms (specifically Tiktok and Instagram) have different effect on academic performance? & \textbf{Z-test} \\
P2 & Does longer sleep duration positively influence mental health among students? & \textbf{Linear Regression} \\
P3 & Does a higher addiction score negatively influence academic performance among students? & \textbf{Linear Regression} \\
P4  & How does social media usage relate to students' mental health across different academic levels? & \textbf{Multiple Linear Regression} \\
P5 & Does higher average social media usage negatively affect average sleep hours per night? & \textbf{Linear Regression} \\
\bottomrule
\end{tabularx}
\end{table*}

\begin{table*}[h]
\centering
\caption{Summary of participants’ time on the implementation task (in minutes), number of interaction turns with ChatGPT, fixed and remaining errors, and the percentage of errors successfully addressed. Participants fixed an average of 63.3\% of the errors.}
\label{tab:field-result}
\resizebox{\textwidth}{!}{%
\begin{tabular}{l l l l l l}
\toprule
\textbf{ID} & \textbf{Time} & \textbf{\# Turns} & \textbf{Fixed Errors} & \textbf{Remaining Errors} & \textbf{\% Errors Fixed} \\
\midrule
P1 & 23.1 & 6  & Only partially privatized & Exceeded privacy budget & 50\%  \\
P2 & 44.3 & 19 & Misused sensitivity, Overly noisy, Exceeded privacy budget & - & 100\% \\
P3 & 45.2 & 14 & Only partially privatized, Exceeded privacy budget & - & 100\% \\
P4 & 25.8 & 8  & - & Misused sensitivity & 0\% \\
P5 & 45.5 & 6  & Overly noisy result, Exceeded privacy budget & Exceeded privacy budget & 67\% \\
\bottomrule
\end{tabular}
}
\end{table*}

\sssec{Results.} Table~\ref{tab:rq-and-analysis} shows the research questions and corresponding analysis methods proposed by the participants. All five participants formulated different questions. Three selected linear regression as their analysis method, one chose multiple linear regression, and one opted for a z-test.

We observed a diverse range of strategies for leveraging LLMs to implement the DP program. P2 and P4 iteratively challenged the ChatGPT-generated code by prompting about potential issues mentioned in \name. In contrast, P1 and P5 reviewed the code themselves and pointed out mistakes. P3 adopted a mixed strategy by identifying some underlying errors on their own and then asking the LLM to perform a final check for all the errors mentioned in \name. On average, participants interacted with ChatGPT 10.6 times (SD = 5.7) and spent 36.8 minutes (SD = 11.3) to complete the implementation task.

Participants were able to generalize the errors to their own cases, despite both the worked and erroneous examples involving differentially private proportions. For example, P1 identified an ``\textit{Only partially privatized}” error and noted, ``\textit{Here you only add noise to true yes counts, not total counts}.'' P3 prompted, ``\textit{Is it possible to have tighter sensitivity estimates?}'' and successfully obtained a more accurate result. Table~\ref{tab:field-result} shows all errors corrected by each participant.

However, fixing one error sometimes introduced other mistakes. For example, P5 initially implemented a differentially private linear regression model by injecting Laplace noise into $X^TX$ and $X^Ty$. They then observed that the results were too noisy, as the slope and intercept fluctuated significantly between runs. Through trial and error, they found that injecting noise directly into the independent and dependent variables ($X$, $y$) produced more stable results. This version did not split the privacy budget between the two releases. Ultimately, three participants had one remaining error in their solutions. On average, participants identified 63.3\% of the errors that appeared.

In the exit survey, participants reported moderate mental demand (Mean = 4.6, SD = 1.7) and low frustration (Mean = 2.2, SD = 1.1). When asked about their strategies for approaching the task, they described how the common issues outlined in \name\ helped guide them through the process. For example, P1 explained, ``\textit{I followed the tutorial to check for common issues like sensitivity or privacy budget allocation}.'' Similarly, P2 noted, ``\textit{First, I followed the tutorial to get an initial answer from ChatGPT, then I went back to ask GPT to solve each possible error}.''

\section{Discussion \& Future Work}

\subsection{Adapting \name\ to Evolving LLMs}
\name\ offers a human-centric approach to support practitioners in verifying and modifying LLM-generated DP code. We designed the worked and erroneous examples around the common mistakes observed in current LLMs and the ways practitioners interact with them. While model performance will improve over time and practitioner-LLM interactions may evolve accordingly, LLMs will still have the chance to provide wrongful information—an innate limitation of the models themselves ~\cite{xu2024hallucination, banerjee2025llms, jones2025ai}. What changes are the common types of errors in the DP code produced by the models. Practitioners will therefore continue to need assistance in validating and correcting LLM-generated DP solutions. By continuously updating worked and erroneous examples to reflect both changes in error patterns and evolving practitioner-LLM behaviors, \name\ can remain an effective scaffold for lowering barriers to implementing DP with LLMs.

\subsection{Training the Next Generation of DP Practitioners}
Most university-level differential privacy (DP) courses focus on mathematical theory, aiming to train future researchers~\cite{cs860kamath, cs3110near, dsc291wang, cs2080Vadhan}. In contrast, few resources offered support for practitioners who implement DP in real-world systems.  The emergence of large language models (LLMs) opens new opportunities: LLMs can now generate DP code with minimal user input, lowering the barrier to adoption. We argue that differentiation—the ability to identify whether a DP implementation is correct or flawed—is a critical skill for practitioners. By training practitioners to be strong differentiators, we can reduce the risk of deploying unsafe or incorrect programs.

\subsection{Applying Cheat Sheets to Other Domain-Specific Tasks}

The idea behind \name\ is that, given DP implementation's structured design space with well-defined checkpoints, we can outline a sequence of steps and guidelines to support novices' verification and correction of LLM-generated code at each stage. These information then allowed us to design both worked and erroneous examples that help practitioners scaffold their implementation of DP with LLMs in ways that resemble experts' approach. The cheat sheet could generalize to other domains where experts rely on structured workflows while novices often lack the expertise to verify or correct LLM mistakes step by step. For example, in the field of machine learning, novices may face challenges such as validating data preprocessing steps, tuning hyperparameters, choosing models, and selecting evaluation metrics ~\cite{skripchuk2022identifying}. A cheat sheet with paired worked and erroneous examples could help them outline a workflow to follow, identify faulty LLM-generated decisions, and align their workflow more closely with expert strategies.

\section{Limitations}

\sssec{Baseline Selection.} The baseline group in our error-identification study received an online tutorial video along with a handout outlining practical DP knowledge. This setup reflects a simplified scenario similar to developers learning DP through online resources. We tried alternative support methods, including handouts alone, free web searches, and other reference materials, but found them less effective. Materials that were too narrow missed essential DP concepts, while overly broad materials required extra effort to identify relevant content. Our baseline emphasizes the DP concepts covered rather than the specific ways they are represented. We acknowledge that our baseline selection represents just one of many possible ways to support self-learning in DP.

\sssec{Generalizability of Error Types.} When identifying the errors introduced by LLMs, we tested only GPT-4o, which was state-of-the-art at the time. As models continue to evolve, the types and frequency of errors may change. Also, we focused on a representative DP scenario: applying additive noise to continuous data to achieve pure DP in a non-interactive setting, a configuration widely adopted in commercial and governmental deployments~\cite{khavkin2025differential}. However, the types of errors may not generalize to other settings, such as interactive analyses where privacy budget accumulates over time. While we observed that many mistakes are recurring, our study does not aim to provide a comprehensive taxonomy of all possible error types.

\sssec{Exclusion of DP Library Usage.} Our study focused on generating DP programs without relying on off-the-shelf libraries~\cite{gaboardi2020programming, berghel2022tumult, holohan2019diffprivlib, pydp}, as few open-source codebases use them, making it difficult for LLMs to reliably produce executable code that incorporates these libraries~\cite{zheng2024well}. This choice limits the scope of our findings, and future work could examine how broader adoption of DP libraries might shift the errors and challenges encountered.

\section{Conclusion}

This paper explores how novices implement DP with LLM assistance. Through a needfinding study (N=9), we identified three key challenges: (1) struggles in setting DP configurations, (2) inability to verify whether LLM-generated programs, and (3) difficulty guiding LLMs toward effective solutions. To address these challenges, we developed \name, an example-based scaffold of expert workflows. It consists of a worked example that step-by-step demonstrates the decision-making points for prompting and verifying LLM code, and erroneous examples that prompt learners to actively reason about choices at each decision point when implementing differential privacy. We evaluated \name\ through a between-subjects study with 24 participants, comparing the 12 who used \name\ to 12 in a baseline group with conventional instructional materials. Participants trained with \name\ identified 66.7\% of errors in LLM-generated programs, compared to only 15.0\% in the baseline group. In the open-ended follow-up study (N=5), participants corrected 63.3\% of mistakes to their self-chosen differentially private data analysis tasks, and two successfully produced correct DP programs.

\bibliographystyle{ACM-Reference-Format}
\bibliography{sample-base}

\appendix

\section{Needfinding Study}

\subsection{Prior DP Knowledge Questions} \label{appendix:need-finding-prior-knowledge}

\begin{enumerate}[noitemsep,topsep=3pt,leftmargin=*]
    \item Have you heard about differential privacy before?
    \begin{enumerate}
        \item Yes
        \item No
    \end{enumerate}

    \item What does a small epsilon value imply?
    \begin{enumerate}
        \item Better accuracy
        \item \textit{\textbf{Strong privacy protection}}
        \item Complete removal of personal data
        \item I'm not sure / I have no idea
    \end{enumerate}
    
    \item Please name three techniques used to achieve differential privacy.

    \item How many years of research experience do you have in differential privacy?
    \begin{enumerate}
        \item Never
        \item Less than 1 year
        \item 1--2 years
        \item 2--3years
        \item More than 3 years
    \end{enumerate}

\end{enumerate}

\subsection{Task Description} \label{appendix:task-description}
Imagine you are working at a software company that collects telemetry data to understand product performance. The telemetry dataset contains the following attributes (see Table ~\ref{tab:needfinding-task_attributes}):

\begin{table}[h!]
\centering
\small
\caption{Attributes of the synthetic telemetry dataset.}
\label{tab:needfinding-task_attributes}
\begin{tabular}{|l l|}
    \hline
    \textbf{Attribute} & \textbf{Description} \\
    \hline
    Product Type & A categorical attribute indicating the type of product. Possible values are: \texttt{A}, \texttt{B}, \texttt{C}, \texttt{D}, \texttt{E}, \texttt{F}, and \texttt{Others}. \\
    \hline
    Event Type & The type of event logged. Possible values are: \texttt{open}, \texttt{close}, \texttt{save}, \texttt{reset}, and \texttt{error}. \\
    \hline
    Time of Event & A timestamp indicating when the event occurred, with values ranging from January 1, 2023, to December 31, 2023. \\
    \hline
    User ID & Anonymized ID of users who produced these events while using each product. \\
    \hline
\end{tabular}
\end{table}

Your task is to write a program to identify \textbf{product types with error rates larger than the average}. These are defined as product types with \textbf{z-scores greater than 0}.

\begin{itemize}[noitemsep,leftmargin=*]
    \item \textbf{Error rate}: number of “error” events / total number of events.
    \item \textbf{Average error rate}: the mean error rate across all product types.
\end{itemize}

The results will be shared with collaborators. Therefore, while you have full access to the raw data, the results must comply with \textbf{differential privacy guarantees} to protect user privacy.

\subsection{Codebook} ~\label{appendix:needfinding-codebook}
See Table~\ref{tab:needfinding-codebook}.

\begin{table*}[h!]
\centering
\footnotesize
\caption{The complete codebook derived from our needfinding study.}
\label{tab:needfinding-codebook}

\begin{tabular}{|p{0.15\textwidth}|p{0.2\textwidth}|p{0.57\textwidth}|}
\hline
\textbf{Theme} & \textbf{Sub-Theme} & \textbf{Code} \\
\hline

\multirow{11}{*}{\parbox{0.15\textwidth}{Stuggles in setting DP configurations}}
& \multirow{4}{*}{\parbox{0.2\textwidth}{Lack of accurate instructions from LLM}}
& Difficulty determining whether dataset size qualifies as “large.” \\ \cline{3-3}
&& Difficulty determining whether dataset qualifies as “sensitive” or “highly sensitive.” \\ \cline{3-3}
&& Uncertain about when to apply Laplace noise versus other types of noise. \\ \cline{3-3}
&& Uncertain about how to set the clipping bound. \\ \cline{2-3}

& \multirow{3}{*}{\parbox{0.2\textwidth}{Lack of dependable feedback from LLM}}
& Cannot decide whether clipping is required when the LLM can provide answers to with and without clipping. \\ \cline{3-3}
&& Difficulty interpreting how to set or adjust $\epsilon$ when the LLM accepts any value. \\ \cline{3-3}
&& Confusion about substituting Laplace with other DP mechanisms when LLM provides no guidance. \\ \cline{2-3}

& \multirow{4}{*}{\parbox{0.2\textwidth}{Lack of explanation to privacy requirements}}
& Parameter explanations are too abstract; need for detailed clarifications. \\ \cline{3-3}
&& Difficulty choosing parameter values without understanding how $\epsilon$ protects data. \\ \cline{3-3}
&& Need for worked examples to guide parameter selection in context. \\ \cline{3-3}
&& Explanations perceived as not useful for implementation or comprehension. \\ \hline

\multirow{14}{*}{\parbox{0.15\textwidth}{Unable to verify LLM generated output}}
& \multirow{5}{*}{Information overload}
& Overwhelmed by too many questions; unsure how to begin. \\ \cline{3-3}
&& Lack of patience to process all new DP concepts. \\ \cline{3-3}
&& Feelings of discouragement about being unable to understand all DP concepts. \\ \cline{3-3}
&& Uncertain about which parameters are most important to prioritize. \\ \cline{2-3}

& \multirow{3}{*}{Confused by program output}
& Uncertain about whether noisy outputs with extreme values (e.g., all 0s or 1s) are meaningful. \\ \cline{3-3}
&& Difficulty understanding noisy z-scores close to true values; uncertainty about privacy protection. \\ \cline{3-3}
&& Confusion about why the MSE vs. epsilon graph is not convex; unclear about expected relationships. \\ \cline{2-3}

& \multirow{4}{*}{\parbox{0.2\textwidth}{Lack of explanation in the generated output}}
& Lack of understanding of how the provided solution protects user privacy. \\ \cline{3-3}
&& Confusion when LLM defaults $\epsilon$ to 1.0 without explanation of reasoning or guidance on how to set the value. \\ \cline{3-3}
&& Uncertainty about why Laplace noise is applied and not other noises. \\ \cline{3-3}
&& Difficulty understanding why smaller $\epsilon$ values yield stronger privacy. \\ \cline{2-3}

& \multirow{2}{*}{Misinterpretation}
& Confusion about meaning of D1 and D2 when only one dataset is provided. \\ \cline{3-3}
&& Concern that adding noise alone is insufficient; desire for formal DP inequality ($\Pr[M(D1) \in S] \leq e^{\epsilon} \cdot \Pr[M(D2) \in S]$) to be satisfied. \\ \hline

\multirow{10}{*}{\parbox{0.15\textwidth}{Difficulty in guiding LLMs to effective solutions}}
& \multirow{5}{*}{Unaware of AI mistakes}
& Misapplied noise addition to event counts instead of error rates, resulting in incorrect sensitivity handling. \\ \cline{3-3}
&& Misinterpretation of sensitivity assumptions (e.g., “sensitivity = 1” incorrectly applied).\\ \cline{3-3}
&& Incorrect clipping strategy: lower-bound clipping with U bound misapplied. \\ \cline{3-3}
&& Misapplication of Laplace noise only to averages, leading to invalid DP analysis. \\ \cline{3-3}
&& Copied LLM code directly into notebook without verification. \\ \cline{2-3}

& \multirow{2}{*}{Inconsistent AI guidance}
& Unaware of LLM’s recapped steps contradicting earlier guidance about noise placement. \\ \cline{3-3}
&& Confused by whether to move Laplace noise addition from error counts to total counts. \\ \cline{2-3}

& \multirow{3}{*}{\parbox{0.2\textwidth}{Lost in back-and-forth exploration}}
& Confusion when switching between code generation and conceptual questions disrupted workflow. \\ \cline{3-3}
&& Back-and-forth coding process with frequent mistakes after several turns of conversations. \\ \cline{3-3}
&& After many instructions, LLM has difficulty following users' instructions consistently. \\ \hline

\end{tabular}
\end{table*}

\section{Prompts Used to Evaluate LLM DP Code Generation} ~\label{appendix:llm-prompts}

\subsection{Average}
Please write a program to compute \textbf{the average number of unique product types each user interacts with throughout the year}.

\noindent The analysis should be done with user-level differential privacy (adding or removing one user). The total privacy budget should be within $\epsilon = 1.0$.

\noindent Your goal is to implement the analysis in a way that satisfies user-level differential privacy ($\epsilon = 1.0$) while preserving as much utility as possible.

\noindent The telemetry dataset is attached. It contains the following columns: \\
- Product Type: A categorical attribute indicating the type of product. Possible values include A, B, C, D, E, F, and Others. \\
- Event Type: The type of event logged. Possible values: open, close, save, reset, and error. \\
- Time of Event: A timestamp indicating when the event occurred. Values range from Jan. 1, 2023, to Dec. 31, 2023. \\
- User ID: Anonymized ID of users who produced these events while using each product.

\subsection{Z-score}
Please write a program to compute \textbf{the z-score of the error rates of each product type}, where the error rate of each product type is defined as the number of error events divided by the total number of events.

\noindent The analysis should be done with user-level differential privacy (adding or removing one user). The total privacy budget should be within $\epsilon = 1.0$.

\noindent Your goal is to implement the analysis in a way that satisfies user-level differential privacy ($\epsilon = 1.0$) while preserving as much utility as possible.

\noindent The telemetry dataset is attached. It contains the following columns: \\
- Product Type: A categorical attribute indicating the type of product. Possible values include A, B, C, D, E, F, and Others. \\
- Event Type: The type of event logged. Possible values: open, close, save, reset, and error. \\
- Time of Event: A timestamp indicating when the event occurred. Values range from Jan. 1, 2023, to Dec. 31, 2023. \\
- User ID: Anonymized ID of users who produced these events while using each product.

\subsection{ANOVA}
Please write a program to \textbf{apply one-way ANOVA to evaluate if the mean number of error events per user is significantly different across product types}. Report the p-value.

\noindent The analysis should be done with user-level differential privacy (adding or removing one user). The total privacy budget should be within $\epsilon = 1.0$.

\noindent Your goal is to implement the analysis in a way that satisfies user-level differential privacy ($\epsilon = 1.0$) while preserving as much utility as possible.

\noindent The telemetry dataset is attached. It contains the following columns: \\
- Product Type: A categorical attribute indicating the type of product. Possible values include A, B, C, D, E, F, and Others. \\
- Event Type: The type of event logged. Possible values: open, close, save, reset, and error. \\
- Time of Event: A timestamp indicating when the event occurred. Values range from Jan. 1, 2023, to Dec. 31, 2023. \\
- User ID: Anonymized ID of users who produced these events while using each product.

\subsection{LLM Self-Check}

\noindent Here are some common mistakes. Please check if your implementation has any of the following issues and fix them.

\vspace{0.8em} 

\noindent\textbf{Mistake 1: Unprivatized data-dependent hyperparameters}

\vspace{0.5em} 

{\centering
\begin{minipage}{\linewidth}
\begin{lstlisting}[language=Python, basicstyle=\ttfamily\small]
### Incorrect code example
sensitivity = len(df)
\end{lstlisting}
\end{minipage}}

\vspace{0.5em} 

{\centering
\begin{minipage}{\linewidth}
\begin{lstlisting}[language=Python, basicstyle=\ttfamily\small]
### Correct code example
sensitivity = 10
\end{lstlisting}
\end{minipage}}

\noindent Set a fixed sensitivity value that doesn’t depend on the data to avoid leaking information.

\vspace{0.5em} 

\noindent\textbf{Mistake 2: Misused sensitivity}

\noindent Assume that we are counting the total number of visits across all visitors with visitor-level differential privacy.

\vspace{0.8em} 

{\centering
\begin{minipage}{\linewidth}
\begin{lstlisting}[language=Python, basicstyle=\ttfamily\small]
### Incorrect code example
sensitivity = 1
\end{lstlisting}
\end{minipage}}

\noindent Since one visitor may visit a restaurant multiple times, setting the sensitivity to 1 doesn’t account for the maximum possible change in the result. A single visitor could cause a larger change in the count than this sensitivity allows.

\vspace{0.5em} 

{\centering
\begin{minipage}{\linewidth}
\begin{lstlisting}[language=Python, basicstyle=\ttfamily\small]
### Correct code example
max_visits_per_visitor = 5

# Clip visits per visitor: randomly select at most `max_visits_per_visitor` visits per visitor
df = df.groupby("VisitorId", group_keys=False).apply(
    lambda x: x.sample(n=min(len(x), max_visits_per_visitor))
).reset_index(drop=True)

# Now each visitor contributes to at most 5 visits
sensitivity = max_visits_per_visitor
\end{lstlisting}
\end{minipage}}

\noindent Clip the number of visits each visitor can contribute to ensure that the sensitivity reflects the maximum possible change. This helps to bound the sensitivity and maintain privacy.

\vspace{0.8em} 

\noindent\textbf{Mistake 3: Only partially privatized}

\vspace{0.5em} 

{\centering
\begin{minipage}{\linewidth}
\begin{lstlisting}[language=Python, basicstyle=\ttfamily\small]
### Incorrect code example
noisy_count = df[df["Time spent"] > 60] + np.random.laplace(0, sensitivity/epsilon)
total_visits = len(df) # Not protected by noise
noisy_ratio = noisy_count / total_visits
\end{lstlisting}
\end{minipage}}

\noindent In this example, the ``total\_visits'' is not protected by noise. Since it’s used to compute the ``noisy\_ratio'', this part of the result could still leak information about the dataset.

\vspace{0.5em}

{\centering
\begin{minipage}{\linewidth}
\begin{lstlisting}[language=Python, basicstyle=\ttfamily\small]
### Correct code example
# Split epsilon between two queries
epsilon_per_query = epsilon / 2
# Compute the ratio with noisy count and noisy total
noisy_count = df[df["Time spent"] > 60] + np.random.laplace(0, count_sensitivity /           epsilon_per_query)
noisy_total = len(df) + np.random.laplace(0, total_sensitivity / epsilon_per_query)
noisy_ratio = noisy_count / noisy_total
\end{lstlisting}
\end{minipage}}

\noindent Split epsilon into two parts: one for computing ``noisy\_count'', the other for ``noisy\_total''. Then use both noisy values to calculate the ratio.

\vspace{0.8em}

\noindent\textbf{Mistake 4: Overly noisy result}

\vspace{0.5em} 

{\centering
\begin{minipage}{\linewidth}
\begin{lstlisting}[language=Python, basicstyle=\ttfamily\small]
### Incorrect code example
sensitivity = 1 # The ratio can change at most 1 when adding or removing one individual
true_ratio = true_count / true_total
noisy_ratio = true_ratio + np.random.laplace(0, sensitivity / epsilon)
\end{lstlisting}
\end{minipage}}

\noindent This approach adds noise based on the sensitivity, which is 1. But the true ratio is between 0 and 1, so the noise is large compared to the value. That makes the result unreliable.

\vspace{0.5em}

{\centering
\begin{minipage}{\linewidth}
\begin{lstlisting}[language=Python, basicstyle=\ttfamily\small]
### Correct code example
# Split epsilon equally between two queries
epsilon_per_query = epsilon / 2
# Release the count and the total with differential privacy
noisy_count = df[df["Time spent"] > 60] + np.random.laplace(0, count_sensitivity /           epsilon_per_query)
noisy_total = len(df) + np.random.laplace(0, total_sensitivity / epsilon_per_query)
noisy_ratio = noisy_count / noisy_total
\end{lstlisting}
\end{minipage}}

\noindent An alternative approach adds noise to the count and the total before computing the ratio. Since both values are large compared to their sensitivities, the noise has less impact and the result is more reliable.

\vspace{0.8em}

\noindent\textbf{Mistake 5: Exceeded privacy budget}

\vspace{0.5em} 

\noindent Assume that we are computing the number of long visits (>= 60 minutes) and short visits (< 60 minutes>) on each day using visitor-level differential privacy, where each visitor is guaranteed to visit the restaurant at most once per day. There are 5 days in total, and the overall epsilon should be within `epsilon`.

\vspace{0.5em}

{\centering
\begin{minipage}{\linewidth}
\begin{lstlisting}[language=Python, basicstyle=\ttfamily\small]
### Incorrect code example
sensitivity = 1
# true_long_visit_count and true_short_visit_count are arrays of length 5
noisy_long_visit_count = true_long_visit_count + np.random.laplace(0, sensitivity /                      epsilon)
noisy_short_visit_count = true_short_visit_count + np.random.laplace(0, sensitivity /                      epsilon)
\end{lstlisting}
\end{minipage}}

\noindent In this example, we are releasing 10 queries, two for each day, with each query using privacy budget of ``epsilon''.

\noindent Let's see how these privacy budgets compose:

\begin{itemize}[noitemsep,leftmargin=*]
    \item Since each visitor is guaranteed to visit at most once per day, their data would only contribute to one of the two queries for that day (either long or short visits, not both). So, on any given day, the privacy cost per visitor is ``max(epsilon, epsilon) = epsilon''.
    \item However, a visitor may visit the restaurant on multiple days, so their data could be used across several daily queries. The privacy budgets accumulate across days. If a visitor comes every day, the total cost becomes ``5 * epsilon''.
\end{itemize}

\noindent Releasing all 10 queries consumes ``5 * epsilon'', which exceeds the overall epsilon constraint of ``epsilon''.

\vspace{0.5em} 

{\centering
\begin{minipage}{\linewidth}
\begin{lstlisting}[language=Python, basicstyle=\ttfamily\small]
### Correct code example
sensitivity = 1
epsilon_per_query = epsilon / 5
noisy_long_visit_count = true_long_visit_count + np.random.laplace(0, sensitivity /                      epsilon_per_query)
noisy_short_visit_count = true_short_visit_count + np.random.laplace(0, sensitivity /                      epsilon_per_query)
\end{lstlisting}
\end{minipage}}

\noindent Perform the queries with privacy budget ``epsilon / 5''.

\section{Prior DP Knowledge Questions in the Error Identification Study}
\label{appendix: verification-priordp}

\begin{enumerate}[noitemsep,topsep=3pt,leftmargin=*]
    \item In differential privacy, which value of the privacy parameter $\epsilon$ provides stronger privacy?
    \begin{enumerate}
        \item \textit{\textbf{$\epsilon$ = 0.1}}
        \item $\epsilon$ = 1.0
        \item I don't know
    \end{enumerate}

    \item Releasing two differentially private statistics on the same dataset, one with $\varepsilon_1$ = 0.1 and the other with $\varepsilon_2$ = 0.5, results in a total privacy loss of:
    \begin{enumerate}
        \item $\epsilon$ = 0.1
        \item $\epsilon$ = 0.5
        \item \textbf{\textit{$\epsilon$ = 0.6}}
        \item $\epsilon$ = 0.05
        \item I don't know
    \end{enumerate}
    
    \item If the mechanism $\mathcal M$ returns a number and satisfies differential privacy with $\epsilon$ = 0.1, does $abs(\mathcal M(x))$ satisfy differential privacy, where $abs$ is the absolute value function?
    \begin{enumerate}
        \item No, not necessarily
        \item \textbf{\textit{Yes, for $\epsilon$ = 0.1}}
        \item Yes, for some $\epsilon$ > 0.1
        \item I don't know
    \end{enumerate}

    \item Which of the following is an advantage of using Differential Privacy?
    \begin{enumerate}
        \item It guarantees complete anonymity of the data subjects
        \item It ensures that the data is completely accurate
        \item \textit{\textbf{It provides a tradeoff between privacy and utility of the data}}
        \item It is a computationally simple method for preserving privacy in large datasets
        \item I don't know
    \end{enumerate}

\end{enumerate}

\section{Programs Used in the Error Identification Study} \label{appendix:verification-programs}

Consider a software telemetry dataset. The dataset contains the following columns:
\begin{itemize}[noitemsep,leftmargin=*]
    \item \textbf{Product Type}: A categorical attribute indicating the type of product. Possible values include A, B, C, D, E, F, and Others.
    \item \textbf{Event Type}: The type of event logged. Possible values: open, close, save, reset, and error
    \item \textbf{Time of Event}: A timestamp indicating when the event occurred. Values range from Jan. 1, 2023, to Dec. 31, 2023.
    \item \textbf{User ID}: Anonymized ID of users who produced these events while using each product.
\end{itemize}
\vspace{0.8em}
\textbf{Average:} The following (see Figure~\ref{fig:dp-code}) presents a Python program to compute the average number of events per product type using user-level pure differential privacy with $\varepsilon = 1.0$. However, the implementation is incorrect. Please point out all the errors and briefly explain why they are incorrect.

\noindent \textit{Errors are: (1) Misused sensitivity, (2) Privatized only part of the computation, and (3) Exceeded privacy budget.}

\begin{figure}[htbp]
\centering
\begin{minipage}{\linewidth}
\begin{lstlisting}[language=Python, basicstyle=\ttfamily\small]
import pandas as pd
import numpy as np
# Parameters
EPSILON = 1.0  # Differential privacy parameter
data_file = "telemetry_data.csv"
# Load the dataset
df = pd.read_csv(data_file)
# Step 1: Count events per user per product type
user_product_counts = df.groupby(
    ['User ID', 'Product Type']
).size().reset_index(name='Event Count')
# Step 2: Aggregate total events per product type
# and number of users per product type
product_totals = user_product_counts.groupby(
    'Product Type'
)['Event Count'].sum()
product_user_counts = user_product_counts.groupby(
    'Product Type'
)['User ID'].nunique()
# Step 3: Add Laplace noise and compute noisy average
def laplace_noise(scale):
    return np.random.laplace(loc=0.0, scale=scale)
sensitivity = 1  # Max impact one user can have on the count
noisy_averages = {}
for product in product_totals.index:
    noisy_total = (
        product_totals[product] + laplace_noise(
        scale=sensitivity / EPSILON
        ))
    count = product_user_counts[product]
    noisy_avg = noisy_total / count if count > 0 else 0
    noisy_averages[product] = noisy_avg
# Step 4: Print results
print(
    "Noisy average number of events per product type 
    (with epsilon = 1.0):"
)
for product, avg in noisy_averages.items():
    print(f"Product {product}: {avg:.2f} events")
\end{lstlisting}
\end{minipage}
\caption{LLM-generated Python code for computing user-level differentially private average. We used this program for the error-identification study.}
\label{fig:dp-code}
\end{figure}

\begin{figure}[htbp]
\centering
\begin{minipage}{1.05\linewidth}
\begin{lstlisting}[language=Python, basicstyle=\ttfamily\small]
import pandas as pd
import numpy as np
from scipy.stats import zscore
# Load the dataset
df = pd.read_csv("telemetry_data.csv")
df = df[['Product Type', 'Event Type', 'User ID']]
# Compute total and error counts per user-product pair
user_product_group = df.groupby(['User ID', 'Product Type'])
user_product_counts = user_product_group[
    'Event Type'
].value_counts().unstack(fill_value=0)
user_product_counts['total'] = user_product_counts.sum(axis=1)
user_product_counts['error'] = user_product_counts.get('error', 0)
# Aggregate counts across all users per product
product_totals = user_product_counts.groupby(level=1).sum()
product_totals['true_error_rate'] = (
    product_totals['error'] / product_totals['total']
)
# Compute per-product sensitivity: 
# max user error count / total events for that product
max_user_errors_per_product = user_product_counts[
    'error'
].groupby(level=1).max()
total_events_per_product = product_totals['total']
sensitivity_per_product = (
    max_user_errors_per_product
    / total_events_per_product
)
# Differential privacy parameters
epsilon_total = 1.0
num_products = 7
epsilon_per_product = epsilon_total / num_products
# Apply Laplace noise to error rates
dp_error_rates = {}
for product in product_totals.index:
    true_rate = product_totals.loc[product, 'true_error_rate']
    sensitivity = sensitivity_per_product.get(product, 0)
    noise = np.random.laplace(
        loc=0.0, scale=sensitivity / epsilon_per_product
    )
    dp_error_rates[product] = min(max(true_rate + noise, 0), 1)
# Compute z-scores
dp_error_series = pd.Series(dp_error_rates)
z_scores = zscore(dp_error_series)
# Combine into a results dataframe
results = pd.DataFrame({
    'True Error Rate': product_totals['true_error_rate'],
    'DP Error Rate': dp_error_series,
    'Z-Score': z_scores
})
\end{lstlisting}
\end{minipage}
\caption{LLM-generated Python code for computing user-level differentially private z-scores. We used this program for the error-identification study.}
\label{fig:dp-error-rate-code}
\end{figure}

\noindent\textbf{Z-score:} The following (see Figure~\ref{fig:dp-error-rate-code}) presents a Python program to compute the z-score of the error rates of each product type using user-level pure differential privacy with $\varepsilon = 1.0$. (The error rates are computed by dividing the number of error events by the total number of events.) However, the implementation is incorrect. Please point out all the errors and briefly explain why they are incorrect.

\noindent \textit{Errors are: (1) Unprivatized data-dependent hyperparameters and (2) Overly noisy results.}

\section{Screenshots of the Interactive LLM-Driven DP Programming Interface} \label{appendix:GUI-screenshots}

\begin{figure}
    \centering
    \includegraphics[width=0.45\linewidth]{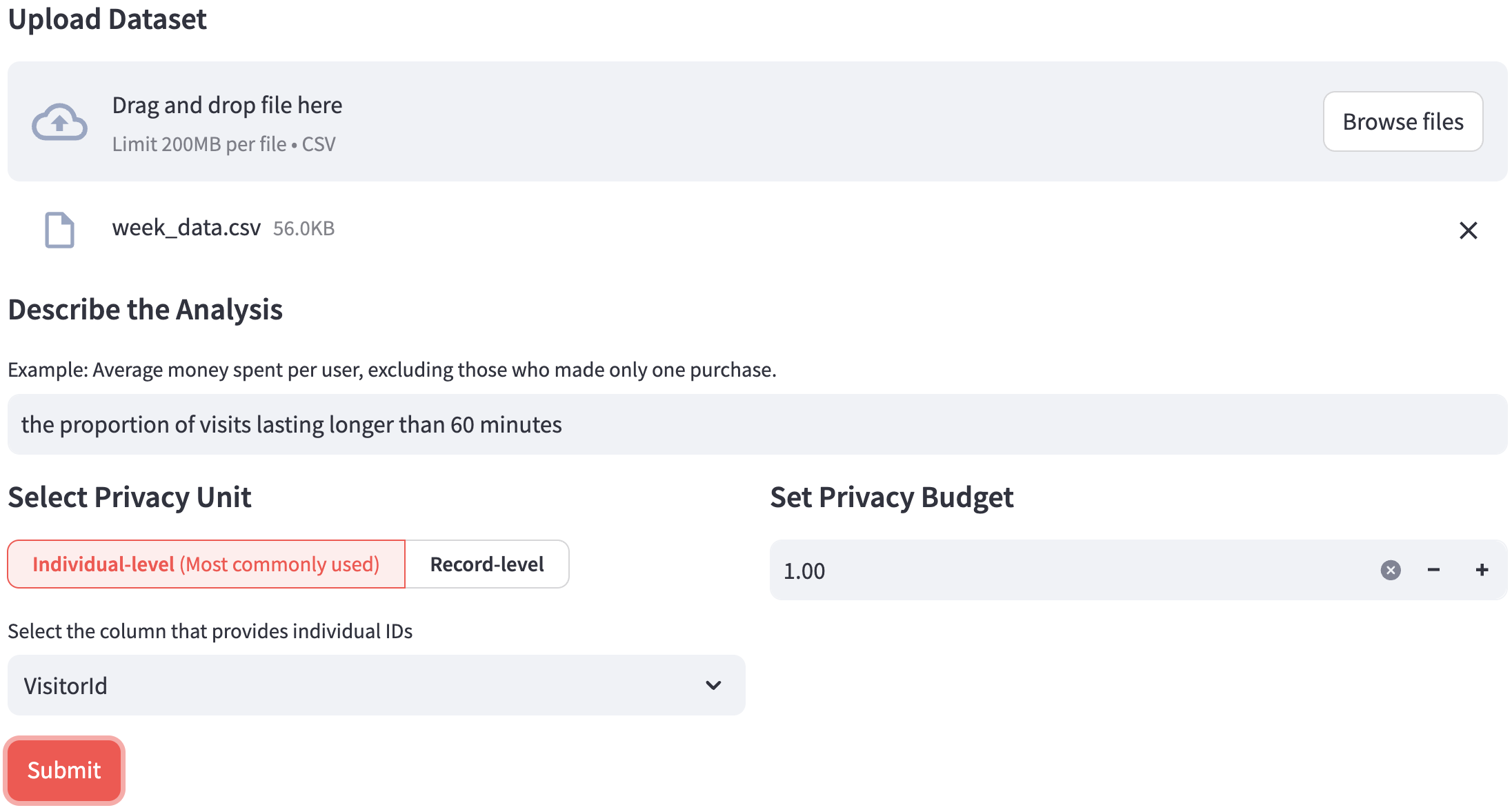}
    \caption{In our initial interactive LLM-driven DP programming interface, users first upload a dataset, specify the query and configure differential privacy parameters.}
    \label{fig:gui-req}
\end{figure}

\begin{figure}
    \centering
    \includegraphics[width=0.45\linewidth]{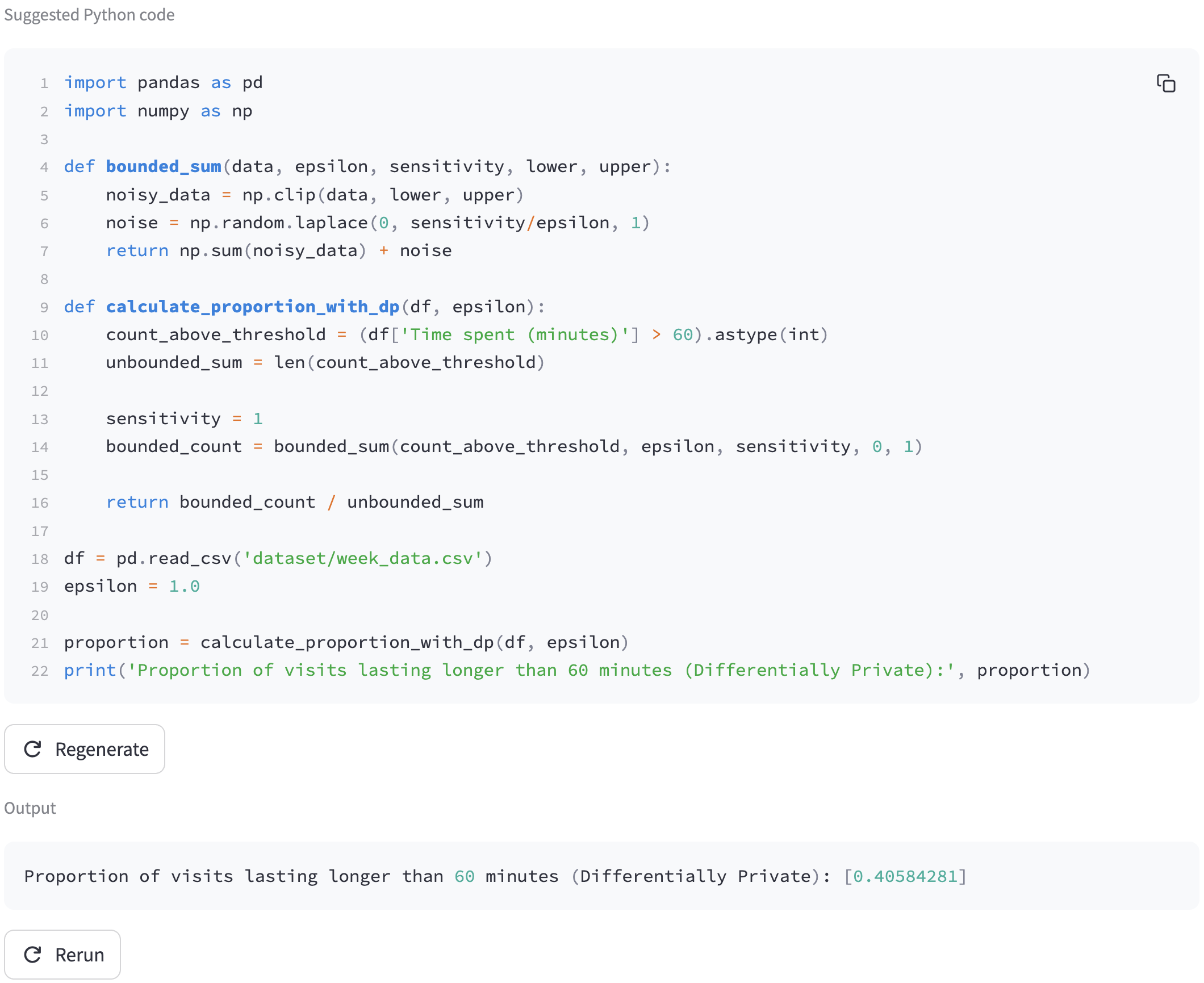}
    \caption{Then, the LLM-powered backbone generates Python code to compute DP solution for the specified query.}
    \label{fig:gui-code}
\end{figure}

\begin{figure}
    \centering
    \includegraphics[width=0.5\linewidth]{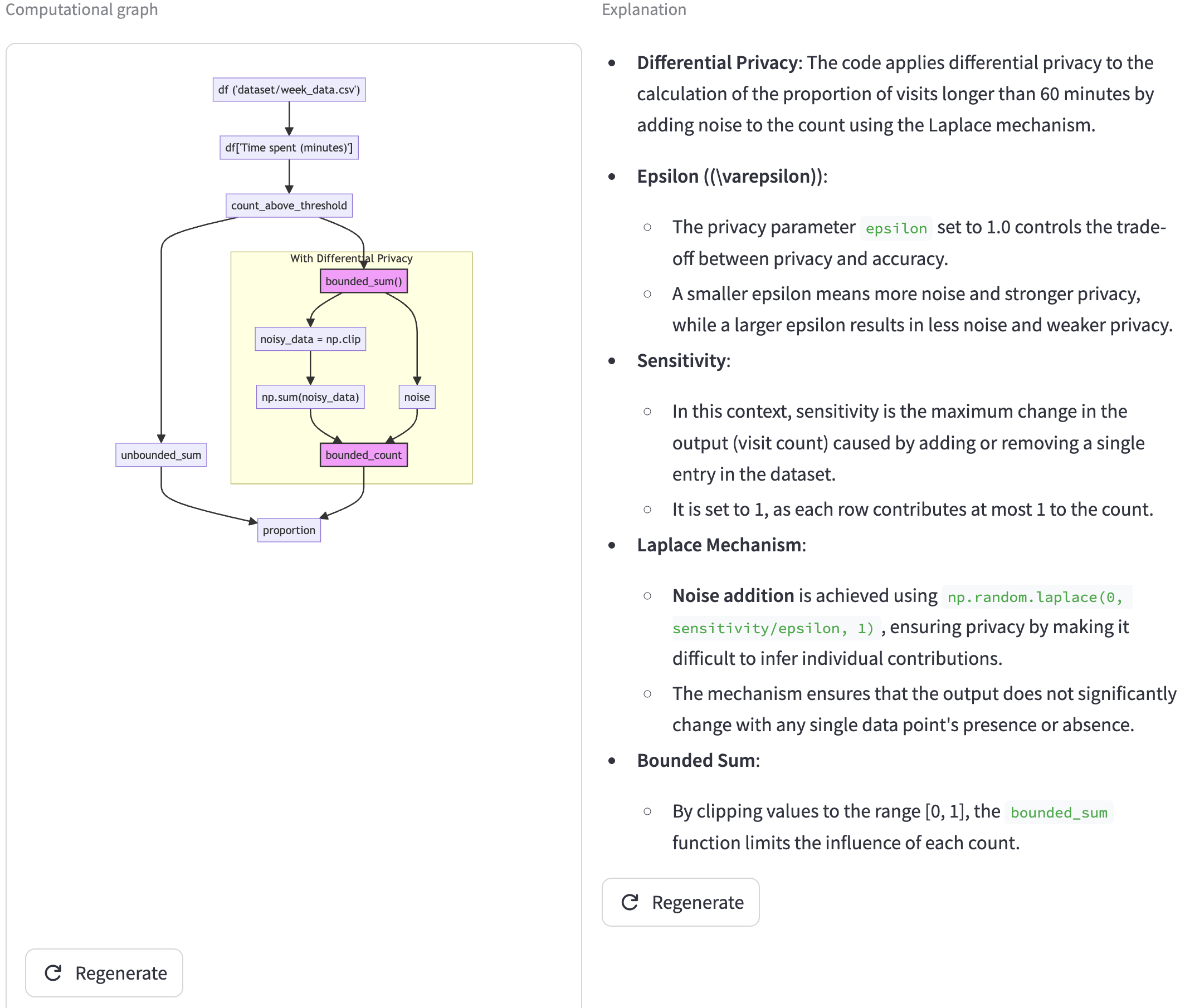}
    \caption{Alongside the program, the LLM also generates a computational graph and explanation to help users understand the code.}
    \label{fig:gui-explanation}
\end{figure}

See Figure~\ref{fig:gui-req},~\ref{fig:gui-code}, and~\ref{fig:gui-explanation}.

\end{document}